\documentclass[a4paper,11pt]{article}
\usepackage[margin=1in]{geometry}
\usepackage{cite}

\usepackage[version=3]{mhchem}
\usepackage{graphicx}
\DeclareGraphicsExtensions{%
    .png,.PNG,%
    .pdf,.PDF,%
    .jpg,.mps,.jpeg,.jbig2,.jb2,.JPG,.JPEG,.JBIG2,.JB2}
\usepackage{color}
\usepackage[nottoc]{tocbibind}
\usepackage[english]{babel}
\usepackage{tabularx}
\usepackage{xparse}
\usepackage{braket}
\usepackage{amsmath}
\usepackage{hyperref}
\usepackage{grfext}
\usepackage{amssymb}
\usepackage{caption}
\usepackage{mathtools}
\usepackage{subcaption}
\usepackage{float}
\usepackage{subcaption}
\usepackage{multirow}
\restylefloat{table}
\usepackage{booktabs}
\usepackage[dvipsnames]{xcolor}
\usepackage{tikz}
\usepackage{pdflscape}
\usepackage{lscape}
\date{\today}
 \normalsize

\newcommand{\bmat}{\left(\begin{array}}
\newcommand{\emat}{\end{array}\right)}
\newcommand{\be}{\begin{equation}}
\newcommand{\ee}{\end{equation}}
\newcommand{\bea}{\begin{eqnarray}}
\newcommand{\eea}{\end{eqnarray}}
\def\ie{{\it i.e.}}

\def\gtwid{\mathrel{\raise.3ex\hbox{$>$\kern-.75em\lower1ex\hbox{$\sim$}}}}
\def\ltwid{\mathrel{\raise.3ex\hbox{$<$\kern-.75em\lower1ex\hbox{$\sim$}}}}
\def\gev{{\rm \, Ge\kern-0.125em V}}
\def\tev{{\rm \, Te\kern-0.125em V}}

\def\lsim{\raise0.3ex\hbox{$\;<$\kern-0.75em\raise-1.1ex\hbox{$\sim\;$}}}
\def\gsim{\raise0.3ex\hbox{$\;>$\kern-0.75em\raise-1.1ex\hbox{$\sim\;$}}}

\def\ie{{\it i.e.,~}}

\def    \be            {\begin{equation}}
\def    \ee            {\end{equation}}
\def    \bea           {\begin{eqnarray}}
\def    \eea           {\end{eqnarray}}

\def\d{\delta}

\def\d{\delta}

 \normalsize

\begin{document}
\renewcommand{\thefootnote}{\fnsymbol{footnote}}
\vspace{.3cm}
\title{\Large\bf Investigating $R_D$ and $R_{D^*}$ anomalies in a Left-Right model with an Inverse Seesaw}

\author
{ \hspace{-2.cm} \it \bf K.
Ezzat$^{1,2}$\thanks{kareemezat@sci.asu.edu.eg},  G.
Faisel$^{3}$\thanks{gaberfaisel@sdu.edu.tr} and S.
Khalil$^{2}$\thanks{skhalil@zewailcity.edu.eg},
 \\\hspace{-2.cm}
\footnotesize$^1$  Department of Mathematics, Faculty of Science, Ain Shams University, Cairo 11566, Egypt.  \\\hspace{-2.cm}
\footnotesize$^2$ Center for Fundamental Physics, Zewail City of Science and
Technology, %\\\hspace{-3.cm}
 \footnotesize  6 October City, Giza 12578, Egypt.  \\\hspace{-2.cm}
\footnotesize$^3$ Department of Physics, Faculty of Arts and Sciences,
S\"uleyman Demirel University, Isparta 32260, Turkey.
}
%\date{\formatdate{2}{1}{2021}}
\date{\today}
\maketitle
%\begin{center}
%\small{\bf Abstract}\\[3mm]
%\end{center}
%%%%%%%%%%%%%%%%
\begin{abstract}
%%%%%%%%%%%%%%%%
We investigate $R_D$ and $R_{D^*}$ anomalies in a low scale
left-right symmetric model based on $SU(3)_C\times SU(2)_L\times
SU(2)_R\times U(1)_{B-L}$ with a simplified Higgs sector
consisting of only one bidoublet and one $SU(2)_R$ doublet. The
Wilson coefficients relevant to the transition $b\to c\tau \nu$
are derived by integrating out the charged Higgs $H^\pm$ boson,
which gives the dominant contributions. We emphasize that the
charged Higgs effects, with the complex right-handed quark mixing
matrix, can account for both $R_D$ and $R_{D^*}$ anomalies
simultaneously, while adhering to a set of significant constraints
including, for instance, ${\rm BR}(B^-_c \to \tau^-
\bar{\nu}_\tau) $ and $B_{s(d)}-\bar B_{s(d)}$ mixing. In relation
to this, we show that the predicted values of the $D^\ast$, $\tau$
longitudinal polarizations and $P_\tau (D)$ can be affected for
the set of the parameters of the model resolving the
$R_{D,D^\ast}$ anomalies.

\end{abstract}
\maketitle
\vskip 0.3cm \hrule \vskip 0.5cm

%**************************************************************
%**************************************************************
\section{Introduction}
%%%%%%%%%%%%%%%%%%%%%%

In flavor physics, the two ratios $R_D$ and $R_{D^*}$ are among
several long-term tensions between the Standard Model (SM) and the
related experimental measurements. These ratios are defined by

\begin{align}
R_{D^*,D} \equiv \frac{\mathcal{B}(B_q\rightarrow \{D^*,D\}
\tau\nu)}{\mathcal{B}(B_q\rightarrow \{D^*,D\} l\nu )}\,,
\end{align}
where $l=e,\mu$. The processes $B_q\rightarrow \{D^*,D\} \ell \nu$
with $\ell=l,\tau$ arise at tree-level from the four fermion
transition $b\rightarrow c\tau\nu$.  Upon combining different
experimental data by BaBar Collaboration
\cite{Lees:2012xj,Lees:2013uzd}, Belle Collaboration
\cite{Huschle:2015rga,Hirose:2016wfn,Hirose:2017dxl,Belle:2019rba}
and LHCb Collaboration
\cite{Aaij:2015yra,Aaij:2017deq,Aaij:2017uff}, the Heavy Flavor
Averaging Group (HFLAV) obtained the average of $R_D$ and
$R_{D^*}$ for 2021 as~\cite{HFLAV:2022pwe},
\begin{align}
R_D &= 0.339 \pm  0.026 \pm  0.014\,, \label{eq:current-data-2}\\
R_{D^*}   &= 0.295 \pm 0.010 \pm 0.010\,.
\label{eq:current-data-1}
\end{align}
Recently, with the announcement of the LHCb collaboration new
results of $R_D$ and $R_{D^*}$ in autumn 2022 using the Run 1 data
sample of a luminosity of $2 fb^{-1}$, the HFLAV obtained a new
averages of $R_D$ and $R_{D^*}$ for 2023 as~\cite{HFLAV:2023},
\begin{align}
R_D &= 0.356 \pm  0.029\,, \label{eq:current-data-22}\\
R_{D^*}   &= 0.284 \pm 0.013 \,. \label{eq:current-data-11}
\end{align}
The  corresponding averages of SM predictions as reported by
HFLAV~\cite{HFLAV:2022pwe} based on the predictions obtained in
Refs. ~\cite{Bigi:2016mdz,Gambino:2019sif,Bordone:2019vic} are
given by
\begin{align}
R_D^{\mathrm{SM}}   &= 0.298 \pm 0.003\,, \\
R_{D^*}^{\mathrm{SM}} &= 0.254 \pm 0.005\,.
\end{align}
Clearly, the listed results of $R_D$ and $R_{D^*}$ in
Eqs.(\ref{eq:current-data-2},\ref{eq:current-data-1}) exceed the
SM predictions given above, by $1.4\sigma$ and $2.9\sigma$
respectively. Moreover, as reported by HFLAV~\cite{HFLAV:2022pwe}, the
difference with the SM predictions reported above, corresponds to
about $3.4\sigma [R_D-R_{D^*}]$ combination. In
Ref.\cite{Iguro:2020cpg}, the authors used all the available
experimental data and took the higher derivative QCDSR constraints
on HQET parameters into account to fit $B_q\rightarrow  \{D^*,D\}$
form factors, and gave the accurate SM prediction of $R_{D^*}$ and
$R_D$. The result of their investigation indicated a $4\sigma$
discrepancy in $R_{D^*}$ and $R_D$. All this may serve as a hint
for a violation of lepton-flavor universality.

In the literature, many theoretical models have been investigated
aiming to explain the $R_D$ and $R_{D^*}$  anomalies. Based on the
remark that the measured ratio $R_{D^*}$ is higher than its value
in the SM, attempts in many models have focused on enhancing the
rate of $b\to c\tau \nu$ transition through extra contributions
arising from the new particles mediating the transition.  This
turns to be much easier than reducing the rate of $b\to c (e,\mu)
\nu$ transitions, given the much more severe restraints on the
couplings of new physics to muons and electrons
\cite{Asadi:2018wea}.

The new particles mediating the $b\to c\tau \nu$ can be spin-0 or
spin-1. In addition, they can either carry baryon and lepton
number (leptoquarks) or be $B/L$ neutral (charged Higgs and $W'$).
As reported in Ref. \cite{Asadi:2018wea}, the existing models can
be classified into three main categories :

\begin{itemize}

\item \textit{Models with charged Higgs}
\cite{Tanaka:2010se,Fajfer:2012jt,Crivellin:2012ye,Celis:2012dk,Iguro:2017ysu,Iguro:2018qzf}.
In these class of models, integrating out the charged Higgs, that
mediates $b\to c\tau \nu$ transition at tree-level results in
scalar-scalar operators contributing to decay rates. The
contributions of these additional operators can be constrained
using the measured $B_c$ lifetime. This leads to the upper limits
${\rm BR}(B_c^{-} \to \tau^{-} \bar{\nu}_\tau) \leq 40\%$
\cite{Celis:2016azn}, ${\rm BR}(B_c^{-} \to \tau^{-}
\bar{\nu}_\tau) \leq 30\%$ \cite{Alonso:2016oyd} and a much
stronger bound ${\rm BR}(B_c^{-} \to \tau^{-} \bar{\nu}_\tau) \leq
10 \%$ was obtained from the LEP data taken at the $Z$ peak
\cite{Akeroyd:2017mhr}. Later on, the bounds using the measured
$B_c$ lifetime had been critically investigated and relaxed upper
limit of $\leq 39 \%$~\cite{Bardhan:2019ljo} and $\leq 60
\%$~\cite{Blanke:2018yud,Blanke:2019qrx} were obtained.  An
analysis of the LHC sensitivity of the charged Higgs that can
explain $R_D$ and $R_{D^*}$ anomalies showed that the bound $
400\, GeV < m_{H^\pm}$ is more stringent than $B_c$ lifetime bound
\cite{Iguro:2018fni}. Based on the finding of
Ref.\cite{Iguro:2018fni} and other investigations carried out
later in Refs.\cite{Iguro:2022uzz, Iguro:2023jju}, the available
charged Higgs mass range  for the explanation of the $R_D$ and
$R_{D^*}$ anomalies within 1\,$\sigma$ range is bounded from the
above as $m_{H^\pm} \leq 400$ GeV. It should be remarked that
$m_{H^\pm} > 400$ GeV is ruled out by the $\tau\nu$ resonance
search at the LHC \cite{Iguro:2018fni,CMS:2018fza}, the low-mass
bottom flavored di-jet search \cite{CMS:2018kcg,ATLAS:2019itm} and
a conventional search for tau sleptons \cite{CMS:2022syk} set
constraint on the available parameter region. Recently, the
analysis  has been extended to cover the mass range $ 180\, GeV <
m_{H^\pm} < 400 \, GeV$ \cite{Iguro:2022uzz,Blanke:2022pjy}.

Experimentally, searches for charged Higgs bosons at high-energy
colliders such as the Large Hadron Collider (LHC) have set
model-dependent limits on their masses. ATLAS and CMS
collaborators have searched for charged Higgs bosons that have
large couplings only to the third generation, looking for multijet
events with one lepton and at least 2 b-jets. They have carried
out dedicated searches only for charged Higgses that dominantly
couple to tb quarks. In addition, there is no dedicated searches
looking for a charged Higgs with sizable couplings to both q b and
t b quarks \cite{Desai:2022zig}.

The production of charged Higgs bosons is influenced by the
particle's mass and its couplings to other particles, which are
determined by the specific beyond the Standard Model (BSM)
scenario being considered. When the charged Higgs boson is heavier
than the top quark, the experimental constraints on its mass
become weaker. Certain regions of the charged Higgs boson mass
parameter space have been excluded by experimental searches, even
for masses higher than the top quark mass. However, these
constraints depend on the couplings to other particles,
particularly top and bottom quarks. We have ensured that our
coupling parameter, $g_{H^- t \bar{b}}$, is below the limits
obtained from previous model-independent analyses in Ref.
\cite{Desai:2022zig}. Therefore, the mass of the charged Higgs
boson can fall within the range of 200 GeV to 1 TeV.

\item \textit{Models with Heavy charged vector bosons}
\cite{He:2012zp,Boucenna:2016qad}. In such class of models,
integrating out $W'$'s yields new contributions to the
vector-vector operators.  Simultaneous explanation of both $R_D$
and $R_{D^{*}}$ with left-handed neutrinos requires non zero
values for the associated new contributions. It should be noted
that, these class of models are subjected to constraints
originating from the presence of an accompanying $Z'$ mediator.
Thus, the vertex $Z'b_Ls_L$ can be inferred from the $W'b_Lc_L$
vertex through the $SU(2)$ invariance. The $Z'b_Ls_L$ term in the
Lagrangian can lead to tree-level flavor-changing neutral currents
(FCNCs) and hence there is a need to some mechanism to subdue the
contributions of this term.  One way to do that is to assume
minimal flavor violation (MFV) as considered in Refs.
\cite{Greljo:2015mma,Faroughy:2016osc}. However, adopting this
choice in general can not suppress $Z'bb$ and $Z'\tau\tau$
vertices. Consequently, LHC direct searches for $Z'\to\tau\tau$
resonances can result in stringent restraints on the related
parameters of the models. However, as shown in Refs.
\cite{Faroughy:2016osc,Crivellin:2017zlb}, selecting unnaturally
high $Z'$ widths helps to avoid these severe constraints.

\item \textit{Models with Leptoquarks}
\cite{Fajfer:2012jt,Tanaka:2012nw}. Either scalar and vector
leptoquarks can be potentially favored as an explanations for the
$R_{D^{(*)}}$ anomaly \cite{Crivellin:2017zlb,Calibbi:2017qbu}.
These models are subjected also to many important constraints
including those from $b\rightarrow s \nu\nu$ restraints
\cite{Crivellin:2017zlb}, searches at the LHC for $\tau\tau$
resonances \cite{Faroughy:2016osc,Crivellin:2017zlb}, and the
measured $B_c$ life-time \cite{Alonso:2016oyd, Akeroyd:2017mhr}.
However, these constraints turn to be weaker compared to the
corresponding ones imposed on the alternative models described
above \cite{Asadi:2018wea}.

\end{itemize}

In this study, we aim to derive the effective Hamiltonian
resulting from the four fermion transition $b\to c\tau \nu$ in the
minimal left-right model with an inverse-seesaw (LRIS). At
tree-level, the transition is mediated by the exchanging of the
charged Higgs boson in the model.
The large neutrino Yukawa, which measures the strength of the charged Higgs-lepton interaction,
and the charged Higgs-quark interaction, which is proportional to right-handed mixing matrix $V^{CKM}_{R}$,
are notable features of this class of Left-right model with an inverse seesaw mechanism that motivate it to solve $R_D$ and $R_{D^*}$.

Having the effective
Hamiltonian, we will proceed to show the dominant Wilson
coefficients contributing to the ratios $R_D$ and $R_{D^*}$.
Consequently, we will display the dependency of the ratios on the
parameters of the model to determine the most relevant ones.
Moreover, we reexamine the relation between the observable
$R_{D^*}$ and ${\rm BR}(B_c \to \tau \bar{\nu}_\tau)$, reported in
Refs.~\cite{Alonso:2016oyd,Akeroyd:2017mhr}, in light of Belle
combination~\cite{Belle:2019rba} and LHCb~\cite{Aaij:2017deq}
experimental results, and the projection at Belle II experiment
with an integrated luminosity of 50 $\rm ab^{-1}$
\cite{Belle-II:2018jsg}. This turns to be very important because
the constraint on ${\rm BR}(B_c \to \tau\bar{\nu}_\tau)$ affects
substantially the contributions from scalar
operators~\cite{Bardhan:2019ljo,Blanke:2018yud,Blanke:2019qrx}.
The $B_{s(d)}-\bar B_{s(d)}$ mixing can lead to possible
constraints on the parameter space related to the the ratios $R_D$
and $R_{D^*}$.  These constraints will be investigated in this
study also. Finally, we will also investigate the possibility of
resolving the anomalies after including the new contributions to
the Wilson coefficients of the total effective Hamiltonian in the
presence of the LRIS model and give predictions of the $D^\ast$
and $\tau$ longitudinal polarizations.

This paper is organized as follows. In Sec.~\ref{model}, we give a
brief review of the Left-Right Model with inverse seesaw. In the
review, we discuss the gauge structure  and the particle content
of the model. We also discuss the fermion interactions with
charged Higgs and with the charged $W,W'$ gauge bosons. These
interactions are required to derive the effective Hamiltonian
governs the processes contributing to $R_D$ and $R_{D^*}$.Then, in
Sec.~\ref{higgs}, we present the effective Hamiltonian describing
$b\to c\tau \nu$ transition in the presence of a general New
Physics (NP) beyond the SM. Particularly, we derive the analytic
expressions of the Wilson coefficients up to one loop level
originating from the charged Higgs mediation in LRIS model under
study in this paper. Based on the Hamiltonian, we show the total
expressions of $R_D$ and $R_{D^*}$. In Sec.~\ref{nums}, we give
our estimation of $R_D$, $R_{D^*}$, the $D^\ast$ and $\tau$
longitudinal polarizations and their dependency on the parameter
space. Finally, in Sec.~\ref{sec:conclusion}, we give our
conclusion.

%%%%%%%%%%%%%%

\section{\label{model}Left-Right Model with inverse seesaw (LRIS)}
The gauge sector of the minimal left-right model with an
inverse-seesaw (LRIS) is based on the symmetry $SU(3)_C\times
SU(2)_L\times SU(2)_R\times U(1)_{B-L}$. The particle content of
the model includes the usual particles in the SM in addition to
extra fermions, scalars and gauge bosons. Regarding the fermion
content of the model, it is the same as its counterpart in the
conventional left-right
models~\cite{Mohapatra:1974gc,Senjanovic:1975rk,Mohapatra:1977mj,Deshpande:1990ip,Aulakh:1998nn,Maiezza:2010ic,Borah:2010zq,Nemevsek:2012iq}.
The scalar sector consists of only one Bidoublet $\phi$ and one doublet $\chi_R$,
which is more minimal and able to circumvent the stringent FCNC constraints placed
on the conventional left-right model. Additionally, we abandon using the left-handed doublet
in order to prevent the severe fine tuning in the VEV of the neutral component of the left-handed doublet scalar, $v_L$,
caused by light neutrino masses. As a result, our model is $SU(2)_L \times SU(2)_R$ gauge invariant but not symmetric in Left-right parity.

The implementation of the IS mechanism for neutrino masses can be
carried out via introducing three SM singlet fermions $S_1$ with
$B-L$ charge $=-2$ and three singlet fermions $S_2$ with $B-L$
charge $=+2$. It should be remarked that, this particular choice
of the $B-L$ charges of the pair $S_{1,2}$ assures that the
$U(1)_{B-L}$ anomaly is free.

The scalar sector of the LRIS model contains  $SU(2)_R$ scalar doublet $\chi_R$ with $B-L$
charge equals -1, and a scalar bidoublet $\phi$ with zero $B-L$ charge. The
extra doublet and bidoublet are essential for breaking the
symmetries of the model upon  having nonvanishing VEVs. We define $\langle
\chi_R\rangle =v_R/\sqrt{2}$ and assume that $v_R$ of an order TeV
to break the right-handed electroweak sector together with $B-L$.
Regarding the VEV of the scalar bidoublet $\phi$, we use the
parametrization $\langle \phi \rangle
=\text{diag}(k_1/\sqrt{2},k_2/\sqrt{2})$ where $k_1=v
s_{\beta},~k_2=v c_{\beta}$ and hence $v^2 =k_1^2 + k_2^2 $ with
$v$ of an order $\mathcal{O}(100)~\text{GeV}$ to break the
electroweak symmetry of the SM. Here and after we use the
definitions $s_x=\sin x,~c_x=\cos x$, $t_x=\tan x$ and $ct_x=\cot
x$.

The scalar-fermion interactions in the LRIS model can be inferred
from the left-right symmetric Yukawa Lagrangian which can be
generally expressed as
\begin{equation}
\mathcal{L}_{\text{Y}}=\sum_{i,j=1}^3 y_{ij}^L \bar{L}_{Li}\phi
L_{Rj} +\tilde{y}_{ij}^L \bar{L}_{Li} \tilde{\phi} L_{Rj} +
y_{ij}^Q \bar{Q}_{Li}\phi Q_{Rj} +\tilde{y}_{ij}^Q \bar{Q}_{Li}
\tilde{\phi} Q_{Rj} +y^s_{ij} \bar{L}_{Ri} \tilde{\chi}_R
S^c_{2j}+H.c., \label{Lag}\end{equation} here $i$ and $j$ are
family indices that run from $1\ldots3$, $y^Q$ and $\tilde y^Q$
represent the quark Yukawa couplings while $y^L$ and $\tilde{y}^L$
stand for the lepton Yukawa couplings. It should be noted that a
$Z_4$ discrete symmetry is implemented in order to forbid a mixing
mass term $M \bar{S^c_1} S_{2}$ and thus preserving the IS
mechanism in a way similar to imposing $Z_2$ symmetry discussed
in~\cite{Khalil:2010iu}. Under this discrete $Z_2$ symmetry,
$S_1$ has a charge $-1$ and all other
fields have a charge $+1$. The fields in Eq.(\ref{Lag}) are
defined as

\be \chi_R=
\begin{pmatrix}\chi_R^+
\\\chi_R^0 \\\end{pmatrix},\,\,\,\, \phi=
\begin{pmatrix}\phi_1^0 &\phi_1^+\\\phi_2^-
&\phi_2^0\\\end{pmatrix},\,\,\,\, Q_{A}=\begin{pmatrix}u_{A}
\\d_{A} \\\end{pmatrix},\,\,\, L_{A}=\begin{pmatrix}\nu_{A} \\e_{A}
\\\end{pmatrix},\,\,\,\, for\,\,\, A=L,R.\ee
The dual bidoublet $\widetilde{\phi}$  and doublet
$\widetilde{\chi}_{R}$  are defined as
$\widetilde{\phi}=\tau_2\phi^*\tau_2$ and
$\widetilde{\chi}_{R}=i\tau_2\chi^*_{R}$ respectively. Clearly,
from the scalar sector of the model and before symmetry breaking,
there are $12$ scalar degrees of freedom: $4$ of $\chi_R$ and $8$
of $\phi$. After symmetry breaking, six scalars out of these $12$
degrees of freedom remain as physical Higgs bosons while the other
degrees of freedom are eaten by the neutral gauge bosons: $Z_\mu$
and $Z'_\mu$ and the charged gauge bosons: $W^\pm_\mu$ and
$W'^\pm_\mu$ to acquire their masses. The physical Higgs bosons
are two  charged Higgs bosons, one pseudoscalar Higgs boson, and
the remaining three are $CP$-even neutral Higgs bosons. For a
detailed discussion about the Higgs sector of the LRIS model we
refer to Ref.\cite{Ezzat:2021bzs}.

We turn now to the neutrino sector of LRIS model. After breaking
the $B-L$ symmetry, one finds that the neutrino
Yukawa interaction terms lead to the following mass
terms\cite{Khalil:2010iu}:%
\be%
{\cal L}_m^{\nu} = m_D \bar{\nu}_L \nu_R + M_R \bar{\nu}^c_R S_2 + h.c.,%
\ee%
where $M_{R}$ and $M_D$ are $3\times 3$
matrices related to $v_R$ and Dirac neutrino masses respectively
via $M_{R}=y^s v_R/\sqrt{2}$ and $M_D=v(y^L s_{\beta}+\tilde{y}^L
c_{\beta})/\sqrt{2}$. In addition one may generate very small Majorana masses for
$S_{1,2}$ fermion through possible non-renormalizable terms.
This tiny mass is required in the standard inverse seesaw mechanism for generating light neutrino masses.
~\cite{Mohapatra:1986aw,Mohapatra:1986bd,GonzalezGarcia:1988rw,Weiland:2013xia}.
Thus, the Lagrangian of
neutrino masses can be expressed as
\be {\cal L}_m^\nu = M_D
\bar{\nu}_L \nu_R + M_R \bar{\nu}^c_R S_2 + \frac{1}{2}\mu_{s_2}
\bar{S^c_2} S_2+H.c.,
 \ee
where $\mu_s \lsim 10^{-6}$ GeV, as it is suppressed by high non-renormalizable scale.
The inverse seesaw mechanism actually depends on small mass scale $\mu_2$, which violates the residual
Lepton number symmetry after breaking the LR symmetry. The light neutrino masses vanish identically in the limit of
$\mu_2 \to 0$, and the lepton number is restored. According to 't Hooft criteria, such a small scale is natural.
We demonstrated that this mass parameter can be generated using a non-renormalizable dimension 7 operator \cite{Khalil:2010iu},
and despite its small size, it plays an important role in generating very small neutrino masses in this mechanism.  In this context,  the
neutrino mass matrix can be written as ${\cal M}_{\nu}
\bar{\psi}^c \psi$
with $\psi=(\nu_L^c ,\nu_R, S_2)$ and ${\cal M}_{\nu}$ is given by %
\begin{equation}
\mathcal{M}_\nu=
\begin{pmatrix}
0 & M_D & 0 \\
M_D^T & 0 & M_{R}\\
0 & M_{R}^T & \mu_{s_2}
\end{pmatrix}.
\label{inverse}
\end{equation}

Note that in order to avoid a possible large mass term $m S_1 S_2$
in the Lagrangian (\ref{Lag}), that would spoil the above
inverse seesaw structure, one assumes that
$L_R$, $\chi_R$, and $S_2$ are even under a $Z_2$-symmetry, while
$S_1$ is an odd particle. Furthermore, similar to $S_2$, the
fermion $S_1$ can acquire a mass via another renormalizable term.
As discussed
in Ref. \cite{El-Zant:2013nta}, $S_1$ can be a natural candidate
for warm dark matter.  It is a kind of sterile neutrino that has
no mixing with active neutrinos and hence  can only interact with
$Z'$ gauge boson. As a consequence, $S_1$ is not subjected to all
constraints imposed on sterile neutrinos due to their mixing with
the active neutrinos. On the the hand, the mass of $S_1$ has to be
of order $O(10 keV)$ to satisfy the combined constraints from
Lyman-$\alpha$ forest data \cite{Zelko:2022tgf} and phase space
arguments for fermionic dark matter \cite{Drewes:2016upu}. This
can be accommodated by choosing $\mu_{s_1}=O(10 keV)$ without any
conflict with the mass parameter $\mu_{s_2}$ contributing to the
neutrino mass matrix given below which is required in our
analysis. Moreover, for $S_1$ being hot dark matter, any related
expected constraints will be set on its mass  $\mu_{s_1}$ and its
couplings to the $Z'$ gauge boson which have no effects on the tau
couplings to neutrinos or neutrino masses or neutrino mixings
needed in this study.

Following the standard procedure, the diagonalization of
$\mathcal{M}_\nu$ results in  the light and heavy neutrino mass
eigen states $\nu_{\ell_i},\nu_{h_j}$ with the  mass eigenvalues
given by:
\begin{align}
\label{nulmass}
m_{\nu_{\ell_i}}&=M_D M_{R}^{-1} \mu_{s_2} (M_{R}^T)^{-1} M_D^T,\quad i=1\ldots3,\\
\label{nuhmass} m_{\nu_{h_j}}^2&=M_{R}^2+M_D^2,\quad j=1\ldots6.
\end{align}
The light neutrino mass matrix in Eq.(\ref{nulmass}) must be
diagonalized by the physical neutrino
mixing matrix $U_{\!M\!N\!S}$ \cite{mns}, {\it i.e.}, %
\be%
U_{\!M\!N\!S}^T m_{\nu_{\ell_i}} U_{\!M\!N\!S} =
m_{\nu_{\ell_i}}^{\rm d} \equiv
{\rm diag}\{m_{\nu_e}, m_{\nu_\mu}, m_{\nu_\tau}\}.%
\ee
Thus, one can easily show that the Dirac neutrino mass
matrix can be defined as :%
\be %
M_D=U_{\!M\!N\!S}\, \sqrt{m_{\nu_{\ell_i}}^{\rm d}}\, R\, \sqrt{\mu^{-1}_{s_2}}\, M_R, %
\label{mD}
\ee %
where $R$ is an arbitrary orthogonal matrix. Clearly, for $M_R$ at
TeV scale and $\mu_{s_2} \ll M_R$ the light neutrino masses can be
of order eV. The $9\times 9$ neutrino mass matrix ${\cal M}_\nu$,
can be diagonalized with the help of the matrix $V$ satisfying
$V^T {\cal
M}_\nu V = {\cal M}_\nu^{\rm diag}$.  The matrix can be expressed as \cite{Dev:2009aw}%
\be V=
\left(%
\begin{array}{cc}
  V_{3\times 3} & V_{3\times6}\\
  V_{6\times 3} & V_{6\times6}  \\
\end{array}%
\right).%
\ee%
As a good approximation, the matrix $V_{3\times3}$ can be given a %
\be%
V_{3\times3} \simeq \left(1-\frac{1}{2} F F^T \right)
U_{\!M\!N\!S} \, . %
\ee%
where  $F = M_D M^{-1}_R$. Generally $F$, as one can see from the
expression of $V_{3\times 3}$, is not unitary matrix. This
unitarity violation, \ie, the deviation from the standard
$U_{\!M\!N\!S}$ matrix, depends on the size of $\frac{1}{2} F F^T$
\cite{Abdallah:2011ew}. The
matrix $V_{3\times6}$ is given by%
\be%
V_{3\times6}=\left({\bf 0}_{3\times3},F \right) V_{6\times6}. %
\label{V36}
\ee %
In addition, $V_{6\times 3}= (V_{3\times 6})^\dagger$. Finally,
the matrix $V_{6\times 6}$ is the one that diagonalize the
$\{\nu_R, S_2\}$ mass matrix.

The neutral scalar fields and their masses can be obtained if one
expands the neutral components of the bidoublet $\phi$ and the
doublet $\chi_{R}$ around their vacua as follows \be \phi_i^0 =
\frac{1}{\sqrt{2}} ( v_i +\phi_i^{0R} +i \phi_i^{0I}), \ee where
$\phi_i=\phi_{1,2},\chi_{R}$ and $v_i=k_{1,2},v_{R}$. In this
case, the symmetric mass matrix of the $CP$-odd Higgs bosons in
the basis $(\phi_1^{0I}, \phi_2^{0I}, \chi_R^{0I})$ is given as
\begin{equation} M_{A}^2=\frac{1}{2}\left(\frac{v_R^2
\alpha_{32}}{c_{2\beta}}-4v^2(2\lambda_2-\lambda_3)\right)
\begin{pmatrix}
c_{\beta}^2 & s_\beta c_\beta & 0 \\
 . & s_{\beta}^2 & 0 \\
 . & . & 0
\end{pmatrix},
\end{equation}
which can be diagonlized by a unitary matrix in the form
\begin{equation}
Z^A=\begin{pmatrix}
 0 & 0 & 1 \\
 -s_{\beta} & c_{\beta} & 0 \\
 c_{\beta} & s_{\beta} & 0
\end{pmatrix},
\end{equation}
leading to $Z^A M_A^2 Z^{AT}=\text{diag}(0,0,m_A^2)$ with the
physical mass of the pseudoscalar boson $A$, $m_{A}^2$, given by
\begin{equation}
m_{A}^2 =\frac{1}{2}\left(\frac{v_R^2}{c_{2\beta}}\alpha_{32}-4v^2(2\lambda_2-\lambda_3)\right).
\end{equation}
Similar to the $CP$-odd Higgs bosons, the  elements of the
$(3\times 3)$ symmetric mass matrix of the $CP$-even Higgs bosons
are given by
\begin{align}
m_{11}&= 2v^2 (\lambda_{1}s_\beta^2+\lambda_{23}c_\beta^2+\lambda_{4} s_{2\beta})+\frac{1}{4}(\frac{1}{c_{2\beta}}+1)\alpha_{32} v_R^2, \\
m_{12}&=m_{21}= v^2 ((\lambda_{1}+\lambda_{23})s_{2\beta}+2\lambda_{4})-\frac{1}{4} \alpha_{32} v_R^2 t_{2\beta}, \\
m_{13}&=m_{31}= v v_R (\alpha_{13} s_{\beta}+\alpha_4 c_{\beta}), \\
m_{22}&= 2v^2 (\lambda_{1}c_\beta^2+\lambda_{23}s_\beta^2+\lambda_{4} s_{2\beta})+\frac{1}{4}(\frac{1}{c_{2\beta}}-1)\alpha_{32} v_R^2, \\
m_{23}&=m_{32}= v v_R (\alpha_{12} c_{\beta}+\alpha_4 s_{\beta}), \\
m_{33}&= 2 \rho_1 v_R^2,
\end{align}
where the potential parameter
$\alpha_{1i}=\alpha_1+\alpha_i,~i=2,3$ and
$\lambda_{23}=2\lambda_2+\lambda_3$.
This matrix can be diagonalized by a unitary transformation matrix $Z^{H}$ such that
 $Z^H M_H^2 Z^{HT}=\text{diag}(m_{H_1}^2,m_{H_2}^2,m_{H_3}^2)$. For more details, we refer to Ref.\cite{Ezzat:2021bzs}.

The process under study can receive dominant contributions from
the charged Higgs mediation at tree level as we will show in the
following. Thus, below, we list the relevant charged Higgs
couplings related to our study following Ref.\cite{Ezzat:2021bzs}.
In the flavor basis $(\phi_1^{\pm},\phi_2^{\pm},\chi_R^{\pm})$,
the charged Higgs bosons symmetric mass matrix takes the form
\begin{equation}\label{mhpm2}
M_{H^\pm}^2=\frac{\alpha_{32}}{2}\begin{pmatrix}
\frac{v_R^2 s_{\beta}^2}{c_{2\beta}} & \frac{v_R^2 s_{2\beta}}{2c_{2\beta}} & -v v_R s_{\beta} \\
 . & \frac{v_R^2 c_{\beta}^2}{c_{2\beta}} & -v v_R c_{\beta} \\
 . & . & v^2 c_{2\beta}
\end{pmatrix}.
\end{equation}
The above matrix can be diagonalized by the unitary matrix,
\begin{equation}
Z^{H^\pm}=\begin{pmatrix}
\frac{v c_{2\beta}}{\sqrt{v^2c_{2\beta}^2+v_R^2s_{\beta}^2}} & 0 & \frac{v_R s_{\beta}}{\sqrt{v^2c_{2\beta}^2+v_R^2s_{\beta}^2}} \\
-\frac{\frac{1}{2}v_R^2
s_{2\beta}}{\sqrt{(v^2c_{2\beta}^2+v_R^2s_{\beta}^2)(v^2c_{2\beta}^2+v_R^2)}}
&
\sqrt{\frac{v^2c_{2\beta}^2+v_R^2s_{\beta}^2}{v^2c_{2\beta}^2+v_R^2}}
& \frac{v v_R
c_{\beta}c_{2\beta}}{\sqrt{(v^2c_{2\beta}^2+v_R^2s_{\beta}^2)(v^2c_{2\beta}^2+v_R^2)}}
\\
-\frac{v_R s_{\beta}}{\sqrt{v^2c_{2\beta}^2+v_R^2}} & -\frac{v_R
c_{\beta}}{\sqrt{v^2c_{2\beta}^2+v_R^2}} & \frac{v
c_{2\beta}}{\sqrt{v^2c_{2\beta}^2+v_R^2}}
\end{pmatrix}.
\end{equation}
The mass eigenstates basis can be obtained using the rotation
$(\phi_1^{\pm},\phi_2^{\pm},\chi_R^{\pm})^T=Z^{H^\pm
T}(G_1^{\pm},G_2^{\pm},H^{\pm})^T$ such that
$Z^{H^\pm}M_{H^\pm}^2Z^{H^\pm T}=\text{diag}(0,0,m_{H^\pm}^2)$.
Here $G_1^{\pm}$ and $G_2^{\pm}$ represent the massless charged
Goldstone bosons and $H^{\pm}$  is a massive physical charged
Higgs boson. The massless Goldstone bosons are eaten by the
charged gauge bosons $W_\mu$ and $W'_\mu$ to acquire their masses
via the familiar Higgs mechanism. On the other hand, the mass of
the charged Higgs boson $H^{\pm}$ is given by:
\begin{equation}
m_{H^{\pm}}^2 = \frac{\alpha_{32}}{2}\left(\frac{
v_R^2}{c_{2\beta}}+v^2c_{2\beta}\right),\label{chHm}
\end{equation}
where $\alpha_{32}=\alpha_3-\alpha_2$ is a potential parameter  Ref.\cite{Ezzat:2021bzs}. Clearly, the charged Higgs
boson mass can be of the order of hundreds~\text{GeV} if we pick
out the values $v_R\sim\mathcal{O}(\text{TeV})$ and
$\alpha_{32}\sim\mathcal{O}(10^{-2})$. It is clear to see that,
the physical charged Higgs boson is a linear combination of the
flavor basis fields $\phi_1^{\pm},\phi_2^{\pm},\chi_R^{\pm}$,
namely given as
\begin{equation}
H^\pm=Z^{H^\pm}_{13}\phi_1^\pm+Z^{H^\pm}_{23}\phi_2^\pm+Z^{H^\pm}_{33}\chi_R^\pm.
\end{equation}
The effective Lagrangian, in the LRIS model, describing the
charged Higgs and the charged gauge bosons $W$ and $W'$ couplings
to quarks and leptons can be expressed as \be \mathcal{L}_{eff}=
\mathcal{L}^{\bar q q' H^\pm}_{eff}+\mathcal{L}^{\bar \nu \ell
H^\pm}_{eff}+\mathcal{L}^{ W,W'}_{eff}\ee The Lagrangians
$\mathcal{L}^{\bar q q' H^\pm}_{eff}$ and $\mathcal{L}^{\bar \nu
\ell H^\pm}_{eff}$ can be obtained from expanding
$\mathcal{L}_{\text{Y}}$, given in Eq.(\ref{Lag}), and rotating
the fields to their corresponding ones in the mass eigenstates
basis. It is direct to obtain
\begin{eqnarray}
\mathcal{L}^{\bar q q' H^\pm}_{eff}&=& \bar{u}_i {\Gamma_{u_i d_j
}^{H^\pm\,LR\,\rm{eff} } }P_R d_j H^\pm\,+ \bar{u}_i {\Gamma_{u_i
d_j }^{H^\pm\,RL\,\rm{eff} } }P_L d_j H^\pm+ h.c.\, ,
 \label{Higgs-vertex}
\end{eqnarray}
where
\begin{eqnarray} {\Gamma_{u_i d_j
}^{H^\pm\,LR\,\rm{eff} } } &=&
 \sum_{a=1}^3 (V^R_{CKM})^*_{j a} y^Q_{i a} Z_{{3 2}}^{H^{\pm}}  + \sum_{a=1}^3
  (V^R_{CKM})^*_{j a} \tilde{y}^Q_{i a} Z_{{3 1}}^{H^{\pm}}
\nonumber\\
{\Gamma_{u_i d_j }^{H^ \pm\,RL\,\rm{eff} } } &=&
\sum_{a=1}^{3}(V^L_{CKM})^*_{j a} y^{Q^*}_{a i} Z_{{3
1}}^{H^{\pm}}  + \sum_{a=1}^{3}(V^L_{CKM})^*_{j a}
\tilde{y}^{Q^*}_{a i}   Z_{{3 2}}^{H^{\pm}}.
 \label{Higgsv1}
\end{eqnarray}
In the LRIS model and after electroweak symmetry breaking, quarks
and charged leptons acquire their masses via Higgs mechanism.
Consequently, we can express the quark Yukawa couplings in terms
of the quark masses and CKM matrices in the left and right sectors
as \bea y^Q &=&\frac{1}{(t_\beta-ct_\beta)}\bigg(\frac{M^{diag}_u
V_{\text{CKM}}^{R \dag}}{v_d} -\frac{V^L_{\text{CKM}} M^{diag}_d}{v_u} \bigg)\nonumber\\
\tilde y^Q &=&\frac{1}{(ct_\beta-t_\beta)}\bigg(\frac{M^{diag}_u
V_{\text{CKM}}^{R \dag}}{v_u} - \frac{V^L_{\text{CKM}} M^{diag}_d
}{v_d} \bigg),\label{yuQ} \eea with $v=246$ GeV, $v_u= \frac{v
s_{\beta}}{\sqrt{2}}$, $v_d= \frac{v c_{\beta}}{\sqrt{2}}$. In the
above equation, $M^{diag}_{u} (M^{diag}_{d})$ is the diagonal up
(down) quark mass matrix, $V^L_{\text{CKM}}$ and
$V^R_{\text{CKM}}$ are the CKM matrices in the left and right
sectors respectively. The mixing matrices for left and right
quarks result in the CKM matrices in the left and right sectors
$V_{\text{CKM}}^{L,R}=V^{u\dag}_{L,R} V^d_{L,R}$. We can choose
the bases where $V^u_L=I$ and as a result, in this basis,
$V_{\text{CKM}}^L=V^d_L$. Turning now to the right sector, we
follow Ref.\cite{Kiers:2002cz}, where the matrix
$V^R_{\text{CKM}}$ can be written as a unitary matrix similar to $V^L_{CKM}$,
but with new mixing angles $\theta^R_{ij}$ and a new Dirac phase $\delta^R$, as follows:
\be
V^R_{\text{CKM}} = K ~
V^L_{\text{CKM}}\left(\theta^R_{12}, \theta^R_{23},\theta^R_{13},
\delta_R \right) ~ \tilde{K}^+, \ee here the diagonal matrices $K$
and $\tilde{K}$ contain five of six non-removable phases in
$V^R_{\text{CKM}}$. It was emphasized that CP violation and FCNC
in the right-handed sector can be under control if the
$V^R_{\text{CKM}}$ is of the form \be V^R_{\text{CKM}}=
\left(\begin{array}{ccc}
1 & 0 & 0 \\
0 & c_{\theta^R_{13}} & s_{\theta^R_{13}} e^{i \alpha} \\
0 & s_{\theta^R_{13}} & -c_{\theta^R_{13}} e^{i\alpha}
\end{array} \label{VRCKMp}\right),
\ee where $c_{\theta^R_{13}}= \cos(\theta^R_{13})$ and
$s_{\theta^R_{13}}= \sin(\theta^R_{13})$ and as we will see that,
in the following, the phase $\alpha$ in the third column plays an
important role in increasing the values of $R_{D^*}$. It is also
possible to have a non-vanishing phase in the second column, which
turns out to be irrelevant and has no effect on the $R_D$ or
$R_{D^*}$ results. Therefore, we set it to zero. We work in the
bases where $V^d_R=I$ and thus $V_{\text{CKM}}^R=V^u_R$, similar
to the left-handed quark sector.

We proceed now to the lepton-neutrinio charged Higgs interactions
in the model under study. These interactions contribute to the
effective Lagrangian $\mathcal{L}^{\bar \nu \ell H^\pm}_{eff}$ and
generally, can be written as
\begin{eqnarray}
\mathcal{L}^{\bar \nu \ell H^\pm}_{eff}&=& \bar\nu_i
{\Gamma_{\nu_i \ell_j }^{H^\pm\,LR\,\rm{eff} } }P_R \ell_j
H^\pm\,+ \bar\nu_i {\Gamma_{\nu_i \ell_j }^{H^\pm\,RL\,\rm{eff} }
}P_L \ell_j H^\pm + h.c.\, ,
 \label{Higgs-vertex}
\end{eqnarray}
with
\begin{eqnarray} {\Gamma_{\nu_i \ell_j }^{H^\pm\,LR\,\rm{eff} } }
&=&\sum_{a=1}^{3} \tilde y^L_{a j} V_{{i a}}  Z_{{3 1}}^{H^{\pm}}
- \sum_{a=1}^{3}y^L_{a j} V_{{i a}}  Z_{{3 2}}^{H^{\pm}}
-\sum_{a=1}^{3}y^{s *}_{j a} V_{{i 6 + a}} Z_{{3 3}}^{H^{\pm}}
\nonumber\\
{\Gamma_{\nu_i \ell_j }^{H^ \pm\,RL\,\rm{eff} } } &=& -
\sum_{a=1}^{3} V^{*}_{{i 3 + a}}  y^{L *}_{j a}   Z_{{3
1}}^{H^{\pm}} + \sum_{a=1}^{3} V^{*}_{{i 3 + a}} \tilde y^{L *}_{j
a}   Z_{{3 2}}^{H^{\pm}}.
 \label{Higgsv2}
\end{eqnarray}
The lepton Yukawa couplings $y^L$ and $\tilde{y}^L$ can be
expressed as
\begin{align}
\label{yukmat} y^L
&=\frac{t_{2\beta}}{\sqrt{2}\,v_u}\bigg(M_\ell-t_\beta M_D \bigg)\,,\\
\tilde{y}^L&=\frac{t_{2\beta}}{\sqrt{2}\,v_u}\bigg(M_D -t_\beta
M_\ell\bigg),
\end{align}
where $M_\ell$ is charged lepton diagonal mass matrix. Finally,
the leptons and quarks gauge interactions related to our processes
at tree-level can be deduced from the effective Lagrangian
$\mathcal{L}^{ W,W'}_{eff}$ given by
\begin{eqnarray}
\mathcal{L}^{ W,W'}_{eff}&=& \frac{1}{\sqrt{2}} g_2 \cos\phi_W
(V_{\text{CKM}}^L)_{ij} W_\mu \bar u_i \gamma_{\mu} P_L d_j -
\frac{1}{\sqrt{2}} g_R \sin\phi_W  (V_{\text{CKM}}^R)^*_{ji} W_\mu
\bar u_i \gamma_{\mu} P_R d_j\nonumber\\&+&\frac{1}{\sqrt{2}} g_2
\sin\phi_W (V_{\text{CKM}}^L)_{ij} W'_\mu \bar u_i \gamma_{\mu}
P_L d_j - \frac{1}{\sqrt{2}} g_R \cos\phi_W
(V_{\text{CKM}}^R)^*_{ji} W'_\mu \bar u_i \gamma_{\mu} P_R  d_j
\nonumber\\&+&\frac{1}{\sqrt{2}} g_2 \cos\phi_W V_{{i j}} W_\mu
\bar \nu_i \gamma_{\mu} P_L e_j -  \frac{1}{\sqrt{2}} g_R
\sin\phi_W  V^{*}_{{i 3 + j}} W_\mu \bar \nu_i \gamma_{\mu} P_R
e_j \nonumber\\&+&\frac{1}{\sqrt{2}} g_2 \sin\phi_W V_{{i j}}
W'_\mu  \bar \nu_i \gamma_{\mu} P_L e_j - \frac{1}{\sqrt{2}} g_R
\cos\phi_W  V^{*}_{{i 3 + j}} W'_\mu \bar \nu_i \gamma_{\mu} P_R
e_j+ h.c.\, ,
 \label{Higgs-vertex}
\end{eqnarray}
where $\phi_W$ is the mixing angle between $W_L$ and $W_R$, which
is of order $10^{-3}$. We have now all the ingredient required for
deriving the effective Hamiltonian contributing to the transition
$b\to c\tau\nu$. In next section, we derive this Hamiltonian and
list the expressions of the Wilson coefficients corresponding to
this Hamiltonian.
\section{\label{higgs} The effective Hamiltonian relevant to the processes in the LRIS}

\begin{figure}[tbhp]
\centering
\includegraphics[width=3cm,height=3cm]{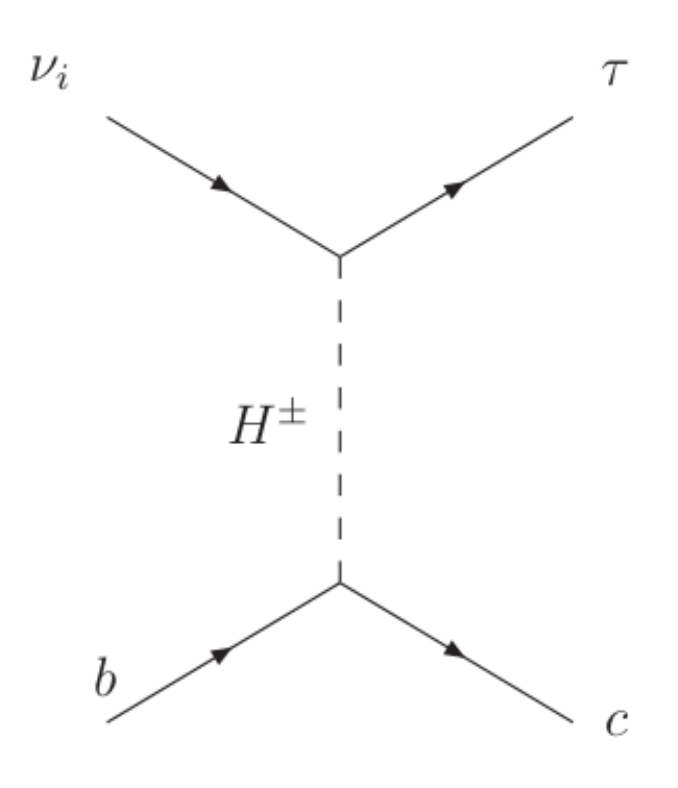}
\hspace{0.2cm}
\includegraphics*[width=3cm,height=3cm]{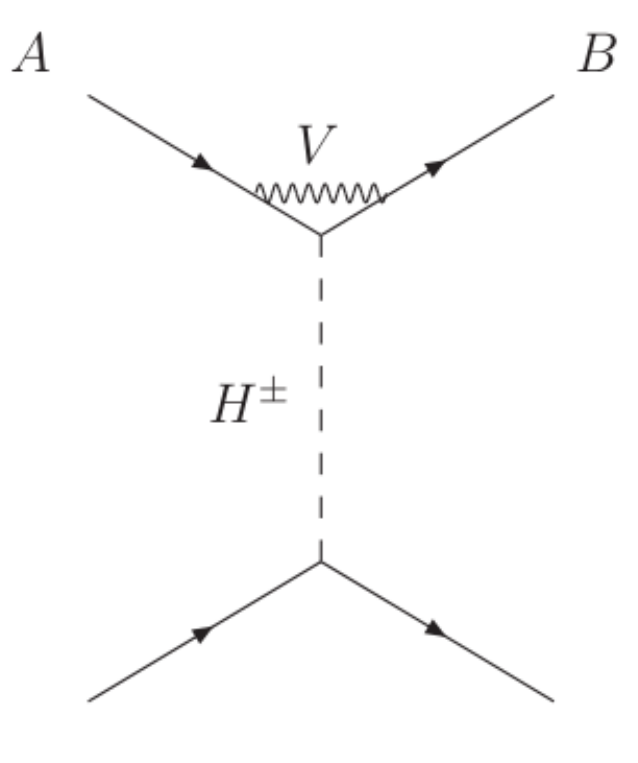}
\hspace{0.2cm}
\includegraphics*[width=3cm,height=3cm]{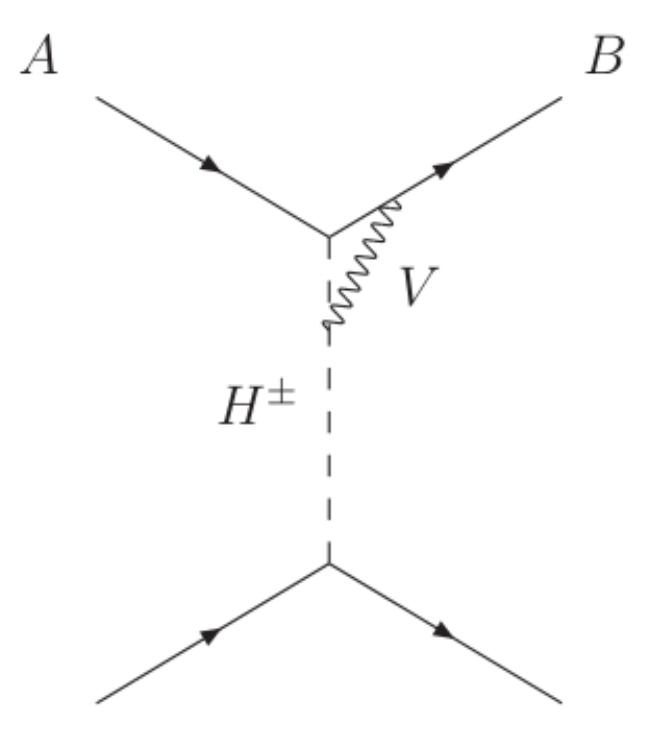}
\hspace{0.2cm}
\includegraphics*[width=3cm,height=3cm]{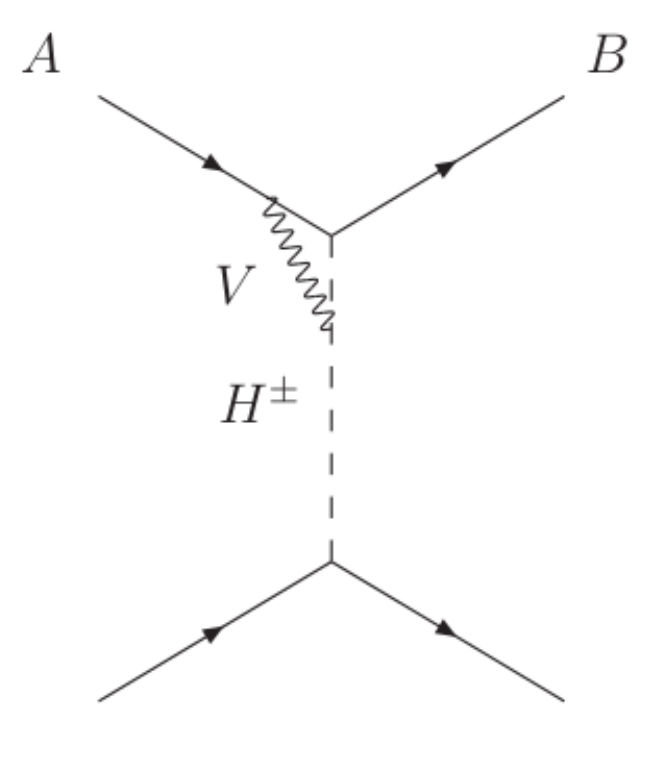}
\vspace{0.7cm}
\includegraphics*[width=3cm,height=3cm]{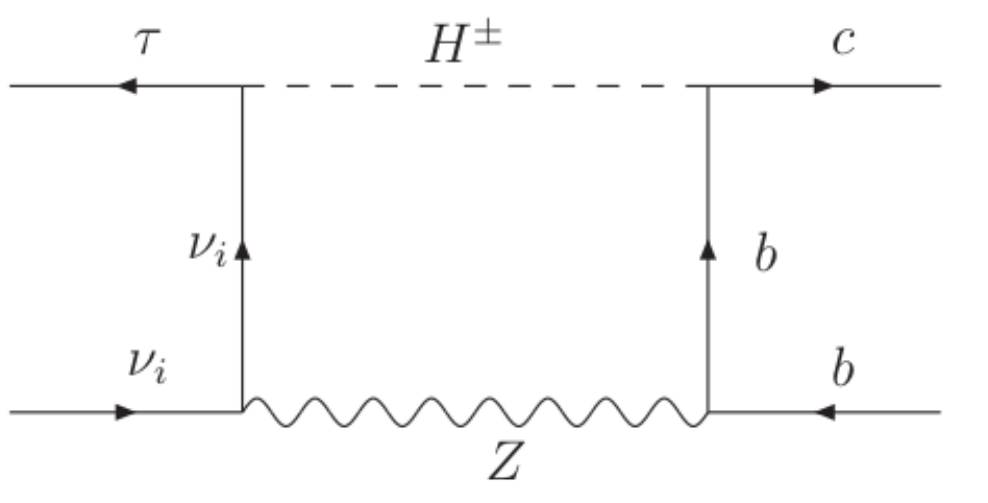}
\hspace{0.2 cm}
\includegraphics[width=3cm,height=3cm]{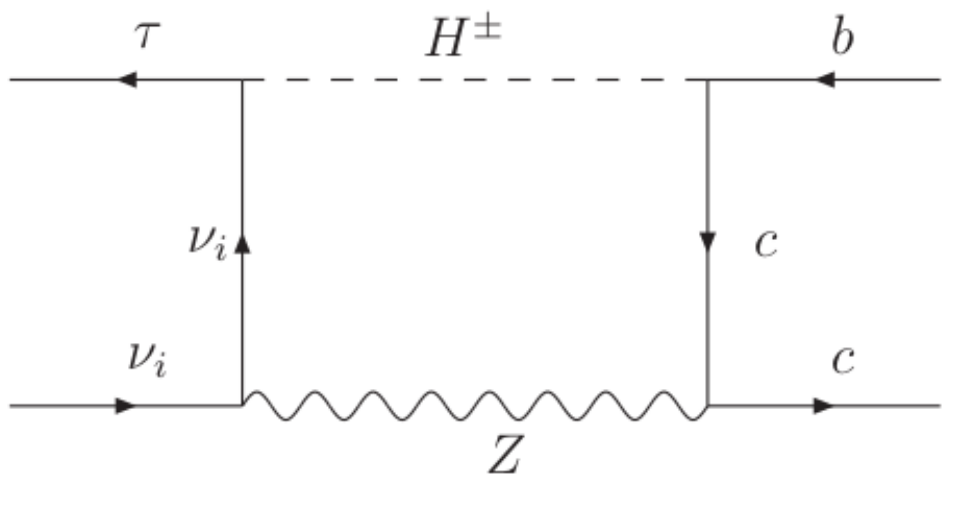}
\hspace{0.2cm}
\includegraphics*[width=3cm,height=3cm]{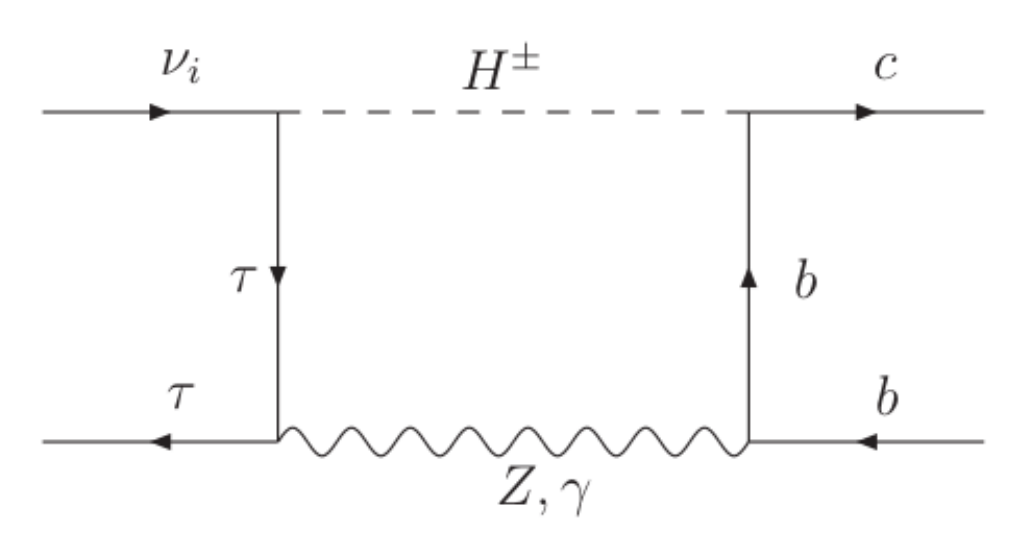}
\hspace{0.2 cm}
\includegraphics*[width=3cm,height=3cm]{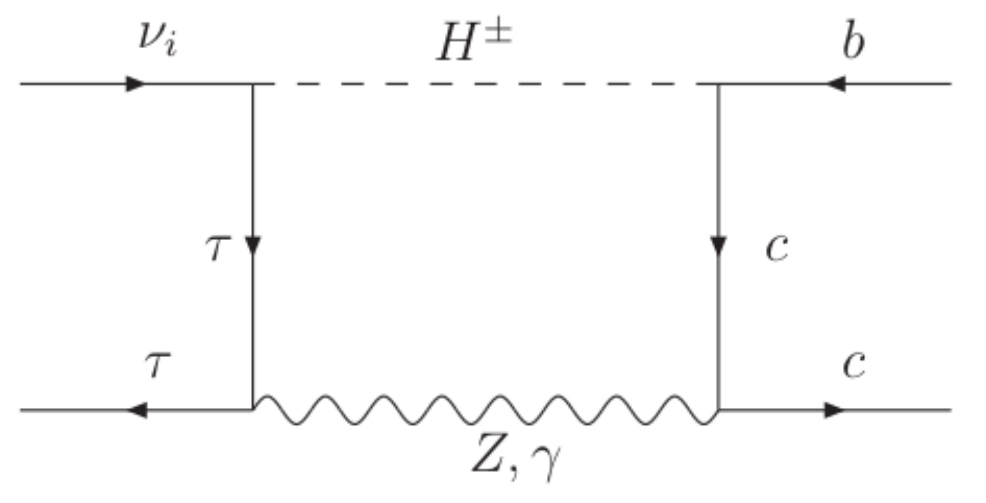}
\medskip
\caption{ Diagrams contributing to ${\mathcal H}_{eff}$, in
Eq.(\ref{Hef}) up to one loop-level due to charged Higgs mediation
with $V$ can be $Z$ or $\gamma$ or both of them depending on the
fermion lines connecting them. In the figure when $A$ and $B$
represent b and c quarks the other two external lines will
represent $\nu_\tau$ and $\tau$. In the case that $A$ and $B$
represent $\nu_\tau$ and $\tau$, the other two external lines
should be understood as representing b and c quarks respectively.
\label{ZBs}}
\end{figure}

In the presence of NP beyond SM, the effective Hamiltonian governs
$\vert \Delta c \vert=1$ B decays transition relevant to our
processes up to one-loop level
derived from the diagrams in Fig.\ref{ZBs}, can be expressed as %

\bea {\mathcal H}_{eff} &=& -\frac{4 G_F}{\sqrt{2}}V_{cb}\bigg(
(1+g_V^{LL})\,(\bar{c} \gamma_\mu P_L b) (\bar \ell_j \gamma^\mu
P_L \nu_i)+ g_V^{RR}(\bar{c} \gamma_\mu P_R b) (\bar \ell_j
\gamma^\mu P_R \nu_i) + g_V^{LR}(\bar{c} \gamma_\mu P_L b) (\bar
\ell_j \gamma^\mu P_R \nu_i)\nonumber\\&+& g_V^{RL} (\bar{c}
\gamma_\mu P_R b) (\bar \ell_j \gamma^\mu P_L
\nu_i)+g_S^{LL}(\bar{c} P_L b) (\bar\ell_j P_L \nu_i) +
g_S^{RR}(\bar{c} P_R b) (\bar\ell_j P_R \nu_i)+g_S^{LR}(\bar{c}
P_L b) (\bar\ell_j P_R \nu_i)\nonumber\\&+&g_S^{RL}(\bar{c} P_R b)
(\bar\ell_j P_L \nu_i)+ g_T^{LL}(\bar{c}\sigma_{\mu \nu }  P_L b)
(\bar \ell_j \sigma^{\mu \nu
}P_L{\nu}_i)+g_T^{RR}(\bar{c}\sigma_{\mu \nu }  P_R b) (\bar
\ell_j\sigma^{\mu \nu } P_R{\nu}_i)\bigg),\label{Hef}\eea where
$V_{cb}$ is the $cb$ Cabibbo-Kobayashi-Maskawa (CKM) matrix
element, $P_{L,R}=\frac{1}{2}(1\mp \gamma_5)$, $\sigma_{\mu\nu}=
\frac{i}{2} [\gamma_\mu, \gamma_\nu]$ and as mentioned in
Ref.\cite{Tanaka:2012nw} the tensor operator with chirality
$(\bar{c}\sigma_{\mu \nu } P_R b) (\bar \ell_j \sigma^{\mu \nu }
P_L{\nu}_i)$ vanishes. In case of $\ell =\tau$ we find that the
Wilson coefficients at the high energy scale  $\mu= m_{H^\pm}$ can
be expressed as

\begin{eqnarray}
g_V^{LL}(\mu) &=&  \frac{m^2_W}{m^2_{W'}} \sin^2\phi_W  Z_{M,{i 3}}^{v} ,\nonumber\\
g_V^{RR}(\mu) &=& \frac{1}{V_{cb}} (\sin^2\phi_W + \frac{m^2_W}{m^2_{W'}} \cos^2\phi_W )(V_{\text{CKM}}^R)^*_{32}  Z^{v,*}_{M,{i 6}}  ,\nonumber\\
g_V^{LR}(\mu) &=& - (1 + \frac{m^2_W}{m^2_{W'}} )  Z^{v,*}_{M,{i 6}}\cos\phi_W \sin \phi_W ,\nonumber\\
g_V^{RL}(\mu) &=& -\frac{1}{V_{cb}} (1 + \frac{m^2_W}{m^2_{W'}} )
(V_{\text{CKM}}^R)^*_{32}  Z_{M,{i3}}^{v}\cos\phi_W \sin \phi_W  ,\nonumber\\
g_S^{LL} (\mu) &=& -\frac{\sqrt{2}}{4 G_F V_{cb} m^2_{H^\pm}} {\Gamma_{u_2 d_3 }^{H^\pm\,RL\,\rm{eff} } }{(\Gamma_{\nu_i \ell_3 }^{H^\pm\,LR\,\rm{eff}})^*  },\nonumber\\
g_S^{RR} (\mu) &=& -\frac{\sqrt{2}}{4 G_F V_{cb}
m^2_{H^\pm}}{\Gamma_{u_2 d_3 }^{H^\pm\,LR\,\rm{eff} }
}{(\Gamma_{\nu_i\ell_3}^{H^\pm\,RL\,\rm{eff} })^* },\nonumber\\
 g_S^{LR} (\mu) &=& -\frac{\sqrt{2}}{4 G_F V_{cb} m^2_{H^\pm}}{\Gamma_{u_2 d_3 }^{H^\pm\,RL\,\rm{eff} } } {(\Gamma_{\nu_i
\ell_3}^{H^\pm\,RL\,\rm{eff} })^* },\nonumber\\
g_S^{RL} (\mu) &=& -\frac{\sqrt{2}}{4 G_F V_{cb} m^2_{H^\pm}}
{\Gamma_{u_2 d_3 }^{H^\pm\,LR\,\rm{eff} } } {(\Gamma_{\nu_i \ell_3
}^{H^\pm\,LR\,\rm{eff} })^* },\nonumber\\
g_T^{LL} (\mu) &=& -\frac{\sqrt{2} \alpha_E}{96\pi G_F V_{cb}
m^2_{H^\pm}} {\Gamma_{u_2 d_3 }^{H^\pm\,RL\,\rm{eff}
}}{(\Gamma_{\nu_i \ell_3 }^{H^\pm\,LR\,\rm{eff} })^* }\bigg(
f(x_c,x_\tau)+\frac{1}{2} f(x_b,x_\tau)\bigg)
+\Delta g^{LL\,Z,Z'}_{{T}},\nonumber\\
g_T^{RR}(\mu) &=& -\frac{\sqrt{2} \alpha_E}{96 \pi G_F
V_{cb}m^2_{H^\pm}} {\Gamma_{u_2 d_3 }^{H^\pm\,LR\,\rm{eff} } }
{(\Gamma_{\nu_i \ell_3 }^{H^\pm\,RL\,\rm{eff} })^* }\bigg(
f(x_c,x_\tau)+\frac{1}{2} f(x_b,x_\tau)\bigg)+\Delta
g^{RR\,Z,Z'}_{ T},\,\label{WiC}
\end{eqnarray}
where $i$ refers to the neutrino flavor, $ x_k
=\frac{m_k^2}{m^2_H}$, $ f(x_i,x_j)=
\frac{1}{(x_i-x_j)}\Big(\frac{x_i}{1-x_i}log
x_i-(x_i\leftrightarrow x_j)\Big)$ and we set the energy scale
$\mu = 1 TeV$. In our  numerical analysis, where all results are
evaluated at the bottom scale $m_b = 4.2 GeV$, the evolution down
of the Wilson coefficients from the scale $\mu = 1 TeV$ to the
bottom scale can be inferred from the renormalization group
evolution (RGE). It should be noted that in Eq.(\ref{WiC}), the
tensor contributions to the Wilson coefficients $ g_T^{LL}(\mu)$
and $g_T^{RR}(\mu)$ are generated from the one-loop diagrams in
the figure. Moreover, in the same equation, we kept only the
dominant contributions to the scalar Wilson coefficients
$g_S^{AB}$, with $A,B$ run over $L,R$, which originate from
tree-level diagrams in Fig.\ref{ZBs}. As we will show below, only
these tree-level scalar contributions will have sizable effects on
the ratios $R_{D,D^\ast}$ and other observables under concern in
this work.

 As can be seen from Eq.(\ref{WiC}), the coefficients $
g_S^{LL}(\mu)$ and $g_S^{RL}(\mu)$ have the same lepton vertex
$(\Gamma_{\nu_i \ell_3 }^{H^\pm\,LR\,\rm{eff}})^*$. Therefore,
they are expected to have different values due to receiving
different contributions of the quark vertices $\Gamma_{u_i d_j
}^{H^ \pm\,RL\,\rm{eff} }$ and $\Gamma_{u_i d_j
}^{H^\pm\,LR\,\rm{eff} }$ expressed in Eq.(\ref{Higgsv1}) in terms
of $y^Q$ and $\tilde y^Q$.  Furthermore, Eq.(\ref{yuQ}) shows that
the contributions from the terms proportional to $V^L_{CKM}$ to
both $y^Q$ and $\tilde y^Q$ are suppressed by the small down quark
masses appear in the matrix $M^{diag}_d$.

 In the processes under consideration,  the corresponding
Wilson coefficient $ g_S^{LL}(\mu)$ ($g_S^{RL}(\mu)$) receives
contributions only from the elements in the second column (row) of
the matrix $M^{diag}_u V_{\text{CKM}}^{R \dag}$, which are present
in both of $y^Q$ and $\tilde y^Q$. With the texture of
$V_{\text{CKM}}^{R}$ given previously, we find that
\be M^{diag}_u V_{\text{CKM}}^{R \dag} = \left(%
\begin{array}{ccc}
  m_u & 0 & 0 \\
  0 & m_c\, c_{\theta^R_{13}} & m_c\, s_{\theta^R_{13}} \\
  0 & m_t\,s_{\theta^R_{13}} e^{-i\alpha} & m_t\,c_{\theta^R_{13}} e^{-i\alpha} \\
\end{array}%
\right).\label{mat} \ee
In this regard, one can show that $R_D$, $R_{D^*}$ and ${\rm
BR}(B^-_c \to \tau^- \bar{\nu}_\tau)$ receive contributions from
the terms proportional to $m_t\,s_{\theta^R_{13}} e^{-i\alpha}$
which originate from $ g_S^{LL}(\mu)$ only and not from
$g_S^{RL}(\mu)$. The other contributions generated from the Wilson
coefficient $g_S^{RL}(\mu)$ are proportional to the charm quark
mass which is small compared to the top quark mass.

Finally in Eq.(\ref{WiC}), the quantities $\Delta
g^{LL(RR)\,Z,Z'}_{ T}$ refer to the suppressed contributions,
comparing to photon ones, originating from $Z$ and $Z'$ mediating
the diagrams. Upon neglecting the small contributions $\Delta
g^{LL,RR\,Z,Z'}_{T }$ we find the following relations

\bea g_T^{LL}(\mu) &=& - \frac{\alpha_E}{24\pi} \bigg(
f(x_c,x_\tau)+\frac{1}{2} f(x_b,x_\tau)\bigg) g_S^{LL}(\mu),
\nonumber\\ g_T^{RR}(\mu) &=& - \frac{\alpha_E}{24\pi} \bigg(
f(x_c,x_\tau)+\frac{1}{2} f(x_b,x_\tau)\bigg) g_S^{RR}(\mu).\eea
For a charged Higgs of a mass $300\,GeV$ we find that
$g_T^{LL}(\mu) \simeq 1.4 \times 10^{-3} g_S^{LL}(\mu)$ and
$g_T^{RR} \simeq 1.4 \times 10^{-3} g_S^{RR}(\mu)$. Clearly the
tensor contributions to the processes under study can be safely
neglected. This is the case also regarding the vector Wilson
coefficients $g_V^{AB}(\mu)$, where $A B$ can be any combination
of the $L$ and $R$ chiralities, as they are suppressed by either
$\frac{m^2_W}{m^2_{W'}}\lesssim 6.4\times 10^{-3}$ for $m_{W'}
\simeq \mathcal{O} (1 TeV)$ or by $\sin \phi_W\simeq \phi_W \simeq
\mathcal{O} (10^{-3})$ or both of them. Consequently, we are left
only with contributions of the scalar Wilson coefficients.

 In the given expressions below, the quantities $g_{S}^{LL}, g_{S}^{RL}, g_{S}^{LR}$ and
$g_{S}^{RR}$ refer to the Wilson coefficients at the bottom scale
$\mu = m_b$. In terms of these quantities, the ratios $R_M$
($M=D,D^\ast)$ are given as
~\cite{Iguro:2018vqb,Asadi:2018wea,Asadi:2018sym}:
\begin{eqnarray}
R_D &=& R_D^{\rm SM} \Big[1+ 1.49 \ \text{Re}(g_{S}^{RL}+
g_{S}^{LL})
+ 1.02 \big( |g_{S}^{RL}+ g_{S}^{LL}|^2 + |g_{S}^{LR}+ g_{S}^{RR}|^2 \big)\Big], \\
R_{D^*} &=&   R_{D^*}^{\rm SM} \Big[1+ 0.11 \
\text{Re}(g_{S}^{RL}- g_{S}^{LL}) + 0.04 \big( |g_{S}^{RL}-
g_{S}^{LL}|^2 + |g_{S}^{LR}- g_{S}^{RR}|^2 \big) \Big].
\end{eqnarray}

It should be noted that in the above expressions of $R_{D,D^\ast}$
we assumed that NP effects are only present in the third
generation of leptons ($\tau, \nu_\tau$). This assumption is
motivated by the absence of deviations from the SM for light
lepton modes $\ell= e$ or $\mu$.

The $D^\ast$ and $\tau$ longitudinal polarizations  depend on the
same Wilson coefficients affecting the $R_{D,D^\ast}$ ratios.
Thus, it is relevant to our investigation to show their predicted
values for the set of the parameters of the model resolving the
$R_{D,D^\ast}$ anomalies. The two observables  have been measured
at Belle experiment. Their expressions can be written as
~\cite{Iguro:2018vqb,Asadi:2018wea,Asadi:2018sym} \small
\begin{eqnarray}
F_L(D^*) &=&  F^{\rm SM}_L(D^*) \ r_{D^\ast}^{-1} \Big[1+ 0.24 \
\text{Re}(g_{S}^{RL}- g_{S}^{LL}) + 0.08 \big( |g_{S}^{RL}-
g_{S}^{LL}|^2
+ |g_{S}^{LR}- g_{S}^{RR}|^2 \big)\Big],\label{FLD_RPV}\\
P_\tau(D) &=&  P^{\rm SM}_\tau(D) \ r_{D}^{-1} \Big[1 +4.65 \
\text{Re}(g_{S}^{RL}+ g_{S}^{LL}) +3.18 \big(
|g_{S}^{RL}+ g_{S}^{LL}|^2 + |g_{S}^{LR}+ g_{S}^{RR}|^2\big)\Big], \\
P_\tau(D^*) &=&  P^{\rm SM}_\tau(D^*) \ r_{D^\ast}^{-1} \Big[1 -
0.22 \ \text{Re}(g_{S}^{RL}- g_{S}^{LL}) - 0.07 \big( |g_{S}^{RL}-
g_{S}^{LL}|^2 + |g_{S}^{LR}- g_{S}^{RR}|^2\big)\Big],
\label{PTAU_RPV}
\end{eqnarray}
\noindent with $r_{D^{(\ast)}} = R_{D^{(*)}} / R^{\rm
SM}_{D^(*)}$. In our analysis we use the measured values of the
$D^\ast$ and $\tau$ longitudinal polarizations reported by Belle
collaborations namely, $F^{Expt}_L(D^\ast) = 0.60 \pm 0.08 \pm
0.035$~\cite{Belle:2019ewo} and $P^{Expt}_\tau(D^\ast) = - 0.38
\pm
0.51^{+0.21}_{-0.16}$~\cite{Hirose:2017dxl,Hirose:2016wfn,Adamczyk:2019wyt}.
On the other hand their SM predictions are estimated as $F^{\rm
SM}_L(D^\ast)= 0.464 \pm 0.010$, $P^{\rm SM}_\tau(D)=0.321 \pm
0.003$ and $P^{\rm SM}_\tau(D^\ast)=-0.496 \pm
0.015$~\cite{Bordone:2019vic}. Although the tau polarization
observable $P_\tau (D)$ is known to be a good discriminator of
scalar contributions originated in many beyond SM physics, (for
instances  the leptoquark (LQ) scenarios \cite{Iguro:2022yzr} )
there is no available experimental measurements of this observable
so far. However, it is important to show its prediction in the
model under study.

The tree-level charged Higgs boson exchange also modifies  the
branching ratio of the tauonic decay $B_c^- \to \tau^-
\bar{\nu}_{\tau}$ as
follows~\cite{Iguro:2018vqb,Asadi:2018wea,Asadi:2018sym} \small
\begin{eqnarray} \label{BRBc_TypeII}
{\rm BR}(B_c^- \to \tau^- \bar{\nu}_{\tau}) &=& {\rm BR}(B_c^- \to
\tau^- \bar{\nu}_{\tau})_{\text{SM}}  \left[ \bigg| 1
 + \frac{m_{B_c}^2}{m_\tau(m_b+m_c)} (g_{S}^{RL}- g_{S}^{LL})\bigg|^2 +
 \bigg|\frac{m_{B_c}^2}{m_\tau(m_b+m_c)} (g_{S}^{LR}- g_{S}^{RR})\bigg|^2\right], \nonumber \\
\end{eqnarray}
\noindent where $m_{B_c}^2/m_\tau(m_b+m_c) = 4.065$ and
 \bea
\label{SM_leptonic} {\rm BR}(B^-_c \to  \tau^-
\bar{\nu}_\tau)_{\rm SM} &=&\tau_{B_c} \dfrac{G_F^2}{8\pi}
|V_{cb}|^2 f_{B_c}^2 m_{B_c} m_{\tau}^2  \Big(1-
\dfrac{m_{\tau}^2}{m_{B_c}^2}\Big)^2 , \eea
where $V_{cb}$ stands for the CKM matrix element, $\tau_{B_c}$ and
$f_{B_c}$  denote the $B^-_c$ meson lifetime and decay constant,
respectively. The SM prediction of ${\rm BR}(B^-_c \to \tau^-
\bar{\nu}_\tau)_{\rm SM} = (2.25 \pm 0.21)\times 10^{-2}$
\cite{Fleischer:2021yjo}. Unfortunately, no direct constraints
from upper bounds on the leptonic $B_c$ branching ratios are
available from the LHC. In view of this, an estimate of a bound on
${\rm BR}(B^-_c \to \tau^- \bar{\nu}_\tau)$ has been derived from
LEP data at the Z peak in Ref. \cite{Akeroyd:2017mhr}. The bound
turns to be strong ${\rm BR}(B_c^{-} \to \tau^{-} \bar{\nu}_\tau)
\leq 10 \%$. Later on, the bounds using the measured $B_c$
lifetime had been critically investigated and relaxed upper limit
of $\leq 39 \%$~\cite{Bardhan:2019ljo} and $\leq 60
\%$~\cite{Blanke:2018yud,Blanke:2019qrx} were obtained. Thus we
will follow Ref. \cite{Fleischer:2021yjo} and  take in our
analysis the bound: ${\rm BR}(B^-_c \to  \tau^- \bar{\nu}_\tau)<
60\% $.

The processes $B_s \to \ell_{A}^{+} \ell_{A}^{-}$, where
$\ell_{A}^{-}$ denotes a charged lepton, can be used to derive
constraints on the parameter space on the model under concern in
this work as we discuss in the following. These processes are
mediated at tree-level by the neutral Higgs ($
H_{k}^{0}=H^{0},h^{0},A^{0}$) exchange. Their branching ratios,
including tree-level neutral Higgs contributions,  can be written
as
\bea &&{\cal{B}}\left[ B^0_s \to \ell_{A}^{+} \ell_{A}^{-}
  \right] = \dfrac{G_{F}^{4} M_{W}^{4} }{8 \pi^{5}}  \,
\sqrt{1-4 x_{i}} \,  M_{ B_s} \,  f_{
B_s}^{2} \,m^{2}_{l_{A}}   \,   \tau_{ B_s}  \nonumber \\
 &\times& \left\{       \left|  \dfrac{M_{B_s}^{2}
  \left(C^{b s}_{P}-C^{' b s}_{P}\right)}{2 m_{l_{A}}\left(m_{s}+m_{b}\right)}
-\left(C^{b s}_{A}-C^{' b s}_{A}\right) \right|^{2} \right. +
\left. \left|  \dfrac{M_{ B_s}^{2} \left(C^{ b s }_{S}-C^{' b
s}_{S}\right)}{2 m_{l_{A}} (m_{s}+m_{b})} \right| ^{2} \times
\left[1-4 x^2_{A}\right] \right\} \,, \label{BRmostgen} \eea
where $x_{i}=\frac{m_{\ell_{i}}}{M_{B_s}}$, $\tau_{B_s}$ and
$f_{B_s}$  stand for the $B^-_s$ meson lifetime and decay
constant, respectively. In the above equation, the neutral Higgs
non-vanishing  Wilson coefficients only include $C^{b s}_{P,S}$
$C^{' b s}_{P,S}$. Their expressions are listed in Eq.(\ref{WCN})
in appendix \ref{Bmumu}. On the other hand and within SM, up to
one loop-level, we have $C^{b s}_{P}=C^{' b s}_{P}=C^{' b
s}_{A}=0$ \cite{Crivellin:2013wna} and
\be { C^{bs}_{A}  = -V^{\star}_{tb}V_{ts} Y\left(\dfrac{
m^{2}_{t}}{ M^{2}_{W}}\right)-V^{\star}_{cb}V_{cs} Y\left(\dfrac{
m^{2}_{c}}{ M^{2}_{W}}\right)\,,   } \label{CASM}
 \ee
The function $Y =\eta_{Y} Y_{0}$ is defined in a way that the NLO
QCD effects are included in $\eta_{Y}=1.0113$
\cite{Buras:2012ru}.The expression of the one loop Inami-Lim
function $Y_{0}$ is given as \cite{Inami:1980fz} \be Y_{0}(x)=
\dfrac{x}{8} \, \left[ \dfrac{4-x}{1-x}+
\dfrac{3\,x}{(1-x)^{2}}\,{\ln (x)} \right] \,.
 \ee
It should be remarked that, the SM Wilson coefficient $
C^{bs}_{A}$ is scale independent as its corresponding effective
operator corresponds to conserved vector current with vanishing
anomalous dimensions \cite{Crivellin:2013wna}.

In our analysis we use the numerical values of the CKM matrix
elements reported in Ref. \cite{UTfit:2022hsi}. Moreover, the
numerical values of the mass and life time of $B_s$ are taken from
Ref.\cite{Workman:2022ynf} and we take the value $f_{B_s}=0.230$
GeV \cite{FlavourLatticeAveragingGroupFLAG:2021npn}. Setting the
neutral Higgs Wilson coefficients to zero in Eq.(\ref{BRmostgen}),
we find that $ B_{SM}(B^0_s\to \mu^{+} \mu^{-}) = 4.1 \times
10^{-9}$ and $ B_{SM}(B^0_s\to \tau^{+} \tau^{-}) = 8.7 \times
10^{-7}$. We note that the SM prediction for the process $
B^0_s\to \mu^{+} \mu^{-}$ obtained here is larger than the well
known one by Misiak et al. (2013) reading $ B_{SM}(B^0_s\to
\mu^{+} \mu^{-}) = (3.65 \pm 0.23) \times 10^{-9}$
\cite{Bobeth:2013uxa}. This can be attributed to the fact that the
authors of Ref. \cite{Bobeth:2013uxa} performed extensive study
and included $ {\cal}O(\alpha_{em})$ and ${\cal}O(\alpha_s^2)$
corrections to the amplitude of the process. Experimentally, from
the non observation of the decay $B^0_s\to \tau^{+} \tau^{-}$, we
have the upper limit $ B(B^0_s\to \tau^{+} \tau^{-}) < 6.8 \times
10^{-3}$ \cite{Workman:2022ynf}. This result allows new physics to
have large contributions to $B(B^0_s\to \tau^{+} \tau^{-})$ and
hence one obtains very loose constraints. This is not the case
regarding the process $B^0_s\to \mu^{+} \mu^{-}$ for which the
experimental measurements $\mathcal{B} \left(B_s \to \mu^+ \mu^-
\right)_{\mathrm{LHCb}} = (3.09^{+0.46+0.15}_{-0.43-0.11})\times
10^{-9}$ \cite{LHCb:2021awg}. We refer to Ref.\cite{HFLAV:2022pwe}
for the results reported by the CMS, ATLAS and CDF collaborations.
Using the $2\sigma$ range of the HFLAV average $\mathcal{B}
\left(B_s \to \mu^+ \mu^- \right)_{\mathrm{HFLAV}} = (3.45\pm
0.29)\times 10^{-9}$ [11] and with the help of
Eq.(\ref{BRmostgen}), we can derive the required constraints on
our parameter space

In the model under consideration in this work, the $B_{s(d)}-\bar
B_{s(d)}$ neutral meson mixing receives new contributions from
tree-level diagrams mediated by the exchange of neutral Higgs
bosons, box diagrams mediated by the charged Higgs only, the $W'$
bosons only, both charged Higgs and $W'$ or $W^\pm$ together and
finally both $W'$ and $W^\pm$ together. In appendix \ref{Bmix}, we
list the set of the operators contributing to the effective
Hamiltonian governing the $B_{s(d)}-\bar B_{s(d)}$ mixing and
their corresponding Wilson coefficients. Following the analysis in
Ref.\cite{DiLuzio:2019jyq}, we can use the reported experimental
and SM values of $\Delta M_{B_{s,d}}$ and $\Delta M_{B_{s,d}}^{\rm
SM}$ respectively to impose the constraints $0.85<{\Delta
M^{SM+H^\pm}_{B_s}}/{\Delta M_{B_s}^{\rm SM}} <1.10$ and
$0.81<{\Delta M^{SM+H^\pm}_{B_d}}/{\Delta M_{B_d}^{\rm SM}} <1.03$
up to $2 \sigma$ level.

The $ b \to s \gamma$ can lead to constraints on the parameter
space of the model under consideration. In our investigation of
these constraints,  we work in leading logarithmic (LL) precision.
Charged Higgs can mediate a loop diagram similar to the one in the
SM but with replacing the charged W bosons with the charged Higgs.
On the other hand, the contributions of the neutral Higgs boson to
$ b \to s \gamma$ are suppressed and thus can be neglected. This
can be explained as the flavor off-diagonal elements in the down
sector can be stringently constrained from the tree-level decays.
Thus, we are left with the contributions originating from charged
Higgs mediating the loop diagram.

In the small $t_\beta$ and $\alpha_{32}$ scheme adopted in this
study, we find that the effective couplings  $\Gamma_{c b
}^{H^\pm\,LR\,\rm{eff} }$ and $\Gamma_{c s }^{H^\pm\,LR\,\rm{eff}
}$ appearing in Eq.(\ref{Higgsv1})  are so tiny, regardless the
values of the parameters $\alpha$ and $\theta_{13}$. This is not
the case if one considers $\Gamma_{c s }^{H^\pm\,RL\,\rm{eff} }$
and $\Gamma_{c b }^{H^\pm\,RL\,\rm{eff} }$ which can be of order
${\cal}O(10^{-2})$ or larger than that. Consequently, in our
analysis, we keep only the dominant contributions to $ b \to s
\gamma$ in the following. Upon, neglecting the operators with mass
dimension higher than six, one obtains the same effective
Hamiltonian as in the case of the SM \cite{Borzumati:1998tg}
\begin{equation}
{\cal{H}}_{eff}^{b \to s \gamma}= -\dfrac{4G_F }{\sqrt{2}}
V_{tb}V_{ts}^\star \sum_i C_{i} \, O_{i}\, .
\end{equation}
In the approximation we adopted above, charged Higgs, propagating
in the loop,  contributes only to the Wilson coefficients
$C^{H^\pm}_7$ and $C^{H^\pm}_8$ corresponding to the operators
\begin{equation}
O_7  = \dfrac{e}{{16\pi^2 }}m_b \bar{s} \sigma ^{\mu \nu } P_R b
F_{\mu \nu }  \quad ; \qquad O_8  = \dfrac{{g_s }}{{16\pi ^2 }}m_b
\bar s\sigma ^{\mu \nu } T^{a} P_R b G^{a}_{\mu \nu }
\end{equation}
The Wilson coefficients $C^{H^\pm}_7$ and $C^{H^\pm}_8$ are given
as
\bea \label{wilson3LO} \nonumber C_{7}^{H^\pm}&\simeq&
\frac{v^2}{\lambda_t}   \sum_{j=1}^{3}~ \Gamma^{RLH^\pm
\star}_{u_j d_2}   \, \Gamma^{RLH^\pm}_{u_j d_3}
\frac{C_{7,YY}^{0}(y_{j})} {m_{u_{j}}^{2}}  \, , \\  \nonumber
 \nonumber C_{8}^{H^\pm}&\simeq & \frac{v^2}{\lambda_t}   \sum_{j=1}^{3}~
 \Gamma^{RLH^\pm \star}_{u_j d_2} \,  \Gamma^{RLH^\pm}_{u_j d_3}
 \frac{C_{8,YY}^{0}(y_{j})} {m_{u_{j}}^{2}} \, , \\  \nonumber
\eea where $y_j=m_{u_j}^2/m_{H^+}^2$ and $\lambda_t=V_{tb} \,
V_{ts}^\star$. The expressions of $C_{7,YY}^{0}$ and
$C_{8,YY}^{0}$ read \cite{Borzumati:1998tg};
\bea \label{c7xyetc} && C_{7,YY}^{0}(y_j) = \frac{y_j}{72} \,
\left[
  \frac{-8y_j^3+3y_j^2+12y_j-7+(18y_j^2-12y_j)\ln y_j}{(y_j-1)^4} \right] \quad ,
\nonumber \\
&& C_{8,YY}^{0}(y_j) = \frac{y_j}{24} \, \left[
\frac{-y_j^3+6y_j^2-3y_j-2-6y_j \ln y_j}{(y_j-1)^4}
  \right] \quad .
\eea From~\cite{Arnan:2019uhr,Misiak:2017woa, Misiak:2015xwa} we
have \bea \dfrac{{{\mathcal B}^{\rm exp}}(b\to s\gamma)}{{\mathcal
B^{\rm SM}}(b\to s\gamma)}-1=- 2.87\,
\left[C^{H^\pm}_7+0.19\,C^{H^\pm}_8\right]=(-0.7\pm 8.2)\times
10^{-2}\,, \eea leading to \cite{Arnan:2019uhr} \bea
|C^{H^\pm}_7+0.19\,C^{H^\pm}_8 | \lesssim 0.06\,\quad
(2\,\sigma)\,. \eea Here, we used $C^{H^\pm}_{7,8}$ at a matching
scale of 1 TeV as input. Again, these constraints are so stringent
that the effect of $C^{H^\pm}_{7,8}$ on the flavour anomalies can
be mostly neglected.

Finally, It is worth to recall that, due to the requirement of
having light neutrino masses we found that the contributions of
the right neutrino sector to the corresponding Wilson coefficients
$ g_S^{RR}(\mu)$ and $g_S^{LR}(\mu)$ are very small and thus can
be safely ignored, leaving us with only $ g_S^{LL}(\mu)$ and
$g_S^{RL}(\mu)$.

\section{Numerical results and analysis \label{nums}}

Having discussed all relevant constraints related to the ratios
$R_{D,D^\ast}$, we are ready now to estimate their predictions in
the model under concern in this investigation. To proceed, we need
first to show, numerically, the RGE running effect of $g^{RL,LL}_S
(\mu)$ from $\mu = 1 TeV$ to $m_b = 4.2 GeV$ scale. According to
the estimation carried out in
Refs.\cite{Gonzalez-Alonso:2017iyc,Iguro:2020keo}, we have

\be \left(%
\begin{array}{c}
    g_S^{RL}\\
   g_S^{LL}  \\
   \end{array}%
\right)=\left(%
\begin{array}{cc}
 1.71 & 0  \\
0 & 1.71  \\
 \end{array}%
\right)\left(%
\begin{array}{c}
g_S^{RL} (\mu = 1 TeV)\\
g_S^{LL} (\mu = 1 TeV)\\
\end{array}%
\right).\ee

\begin{figure}[tbhp]
\includegraphics[width=7cm,height=6cm]{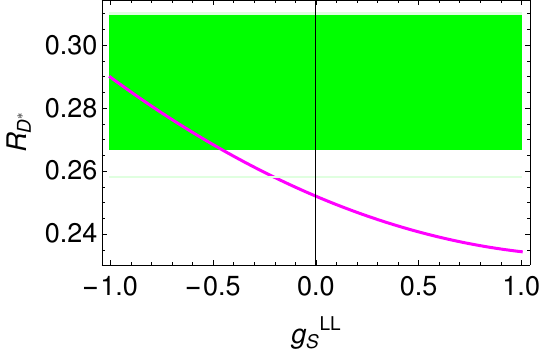}
\hspace{0.2cm}
\includegraphics[width=7cm,height=6cm]{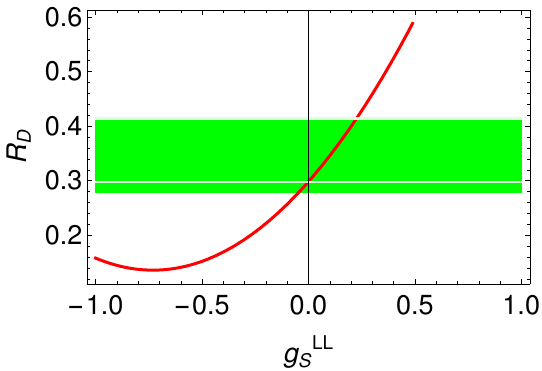}
\medskip
\caption{ Left (right) $R_{D^*}$ ($R_D$) variation with
$g_{S}^{LL}$ where the shaded green regions represent the allowed
$2\sigma$ region of their experimental values. \label{RegsLL}}
\end{figure}

Furthermore, we have checked that $g_S^{RL}$ is about one order of
magnitude smaller than $ g_S^{LL}$; thus for real $ g_S^{LL}$
({\it i.e.}, $ Re(g_S^{LL}) =  g_S^{LL}$), one finds
\begin{eqnarray}
R_D &\simeq& R_D^{\rm SM} \Big[1+ 1.49 g_{S}^{LL}+ 1.02  |g_{S}^{LL}|^2 \Big], \\
R_{D^*} &\simeq&   R_{D^*}^{\rm SM} \Big[1- 0.11 g_{S}^{LL} + 0.04
|g_{S}^{LL}|^2 \Big].
\end{eqnarray}
This expression clearly shows that enhancing the values of
$R_{D^*}$ to be in the range of the given experimental results,
while keeping the limit $ -1\lsim g_S^{LL} \lsim 1$ in mind is
possible for a range of negative values of $g_S^{LL}$ as can be
seen from the left plot in Fig.\ref{RegsLL}. However, as can be
remarked from the right plot in Fig.\ref{RegsLL}, these negative
values reduce $R_D$ below their allowed $2\sigma$ region of the
experimental results shown by the shaded green region in the plot.
Clearly, we deduce that the phase $\alpha$ of the mixing matrix
$V^R_{CKM}$ is crucial to solve the concerned anomalies for the
processes under consideration.

The attainable values of the Wilson coefficients $ g_S^{LL}$ and
$g_S^{RL}$ are affected by the charged Higgs mass. As a result, to
enhance these coefficients and thus the values of $R_D$ and
$R_{D^*}$ while adhering to direct search constraints, the charged
Higgs masses should not be too heavy, namely of the order of
hundreds GeV. According to Eq.(\ref{chHm}), this can be
accomplished by considering $v_R\sim\mathcal{O}(\text{TeV})$,
$\alpha_{32}\sim\mathcal{O}(10^{-2})$ and $t_\beta$ less than one.
Remarkably, from the pre factor $1/(t_\beta-ct_\beta)$ in
Eq.(\ref{yuQ}), it is direct to see that small $t_\beta$ values
can enhance also the quark Yukawa couplings and hence together
with the angle $\theta^R_{13}$ and the complex phase $\alpha$ of
the right-quark mixing $V^R_{CKM}$, defined in Eq.(\ref{VRCKMp}),
play crucial roles in increasing the values of $R_D$ and
$R_{D^*}$, and allow them to take values that are compatible with
the limit of the experiments at the same time.  In our scan of the
parameter space we take $\alpha_{32} \in [0.007,0.016]$, $t_\beta
\in [0.15,0.25]$, $c_{\theta^R_{13}} \in [-1,1]$, $\alpha \in
[0,\pi]$ and $v_R=6400$ GeV. With this in hand, we show below our
results corresponding to the scanned points in the parameter space
respecting all the bounds discussed in the previous section. It
should be noted that, in these results we only select the points
that lead to values of $F_L(D^*)$ and $P_{\tau}(D^*)$ within their
$2\sigma$ range of the corresponding experimental results.
Moreover, we have checked that, for these parameters in the chosen
ranges and values, the contribution of the $g_S^{RL}$ are
irrelevant and can be safely neglected. This confirms our previous
conclusion that only the Wilson coefficient $g_S^{LL}$ plays the
major role through the terms proportional to
$m_t\,s_{\theta^R_{13}} e^{-i\alpha}$. In the following, we
present a set of elucidative plots of $R_D$ and $R_{D^*}$ versus
some selected relevant parameters of the model. It should be noted
that these plots were generated through a random scan and
therefore do not reflect any specific correlations with the chosen
parameter.

\begin{figure}[ht]
\centering
\includegraphics[width=9.5cm,height=5.5cm]{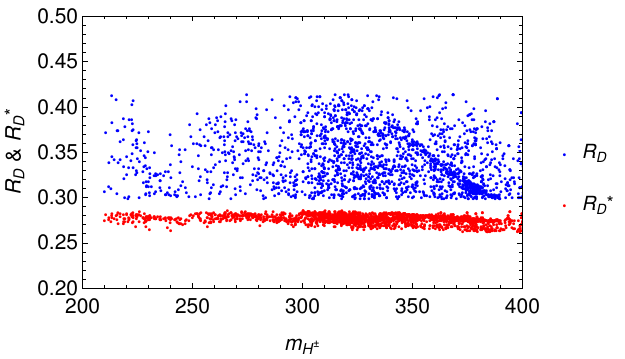}
\caption{$R_D$ and $R_{D^*}$ as function of the charged Higgs mass
where the parameters as stated in the text are chosen as follows:
$c_{\theta^R_{13}} \in [-1,1]$, $\alpha \in [0,\pi]$, $\alpha_{32}
\in [0.00166,0.00716]$ and $t_\beta \in [0.15,0.25]$.} \label{charged
Higgs}
\end{figure}

\begin{figure}[ht]
\includegraphics[width=8.5cm,height=5.5cm]{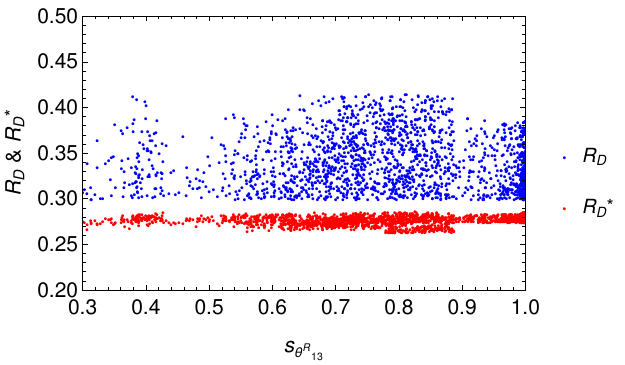}
\hspace{0.2cm}
\includegraphics[width=8.5cm,height=5.5cm]{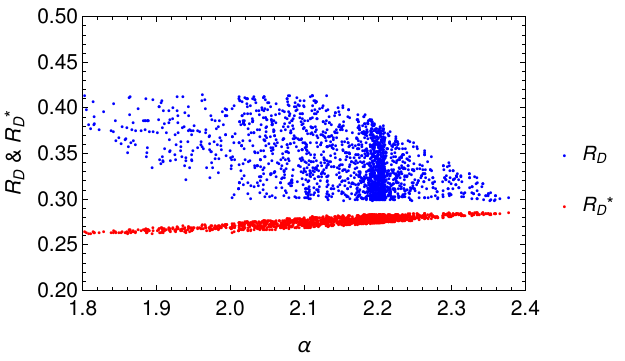}

\caption{$R_D$ and $R_{D^*}$ as function of the sin of  mixing
angle $\theta^R_{13}$ left and the phase $\alpha$ of the
$V^R_{CKM}$ matrix right and other parameters are fixed as in the
previous figure.} \label{arg VR}
\end{figure}

\begin{figure}[tbhp]
\centering
\includegraphics[width=7cm,height=7cm]{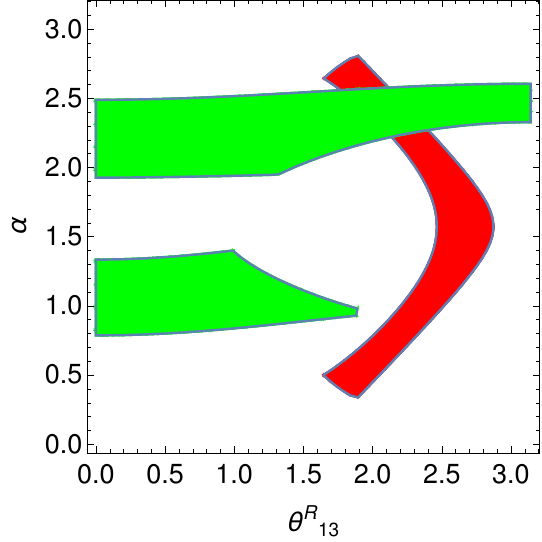}
\hspace{0.2cm}
\medskip
\caption{ Allowed region in the $(\theta^R_{13},\alpha)$ plane by
all considered constraints in green color. The red region
satisfies the $2\sigma$ experimental results of $R_{D^*}$ and
$R_D$ together for $t_\beta\simeq 0.2074$, $\alpha_{32} = 0.00416$
which result in $m_{H^\pm} = 305$ GeV and the other parameters are
fixed as before. \label{ZBs11}}
\end{figure}

\begin{figure}[tbhp]
\centering
\includegraphics[width=8.5cm,height=6.5cm]{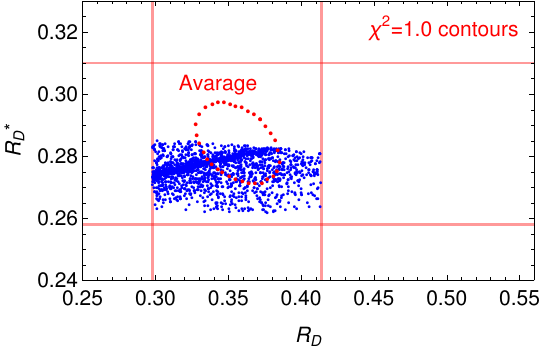}
\medskip
\caption{ The correlation between  $R_D$ and $R_{D^*}$ for the
same set of parameter space considered in Fig. \ref{charged
Higgs}. The  dashed red line represents the $\Delta \chi^2 = 1.0$
contour, which is consistent with the 2023 latest results from
HFLAV \cite{HFLAV:2023}.\label{correlation}}
\end{figure}

In Fig. \ref{charged Higgs}, we display the variation of $R_D$ and
$R_{D^*}$ with the charged Higgs mass. As can be seen from the
figure, it is possible to account for the experimental results of
$R_D$ and $R_{D^*}$ within $2\sigma$ range, while respecting the
the mentioned constraints in the previous section  and the
$2\sigma$ range of $F_L(D^*)$ and $P_{\tau}(D^*)$ mentioned in the
previous section, with charged Higgs masses $m_{H^\pm}$ can be
chosen of order 300 GeV. On the other hand, the dependence of
$R_D$ and $R_{D^*}$ on the sin of the mixing angle $\theta^R_{13}$
and the phase $\alpha$ of the $V^R_{CKM}$ matrix is depicted in
Fig. \ref{arg VR}. It is clear from left plot in the figure that,
moderate and large values of $s_{\theta^R_{13}}$ are preferable to
satisfy the experimental results of $R_D$ and $R_{D^*}$ within
$2\sigma$ range while respecting the bounds and the requirements
considered in the scan. On the other hand, regarding the phase
$\alpha$, large phases are favored as can be seen from the right
plot in the figure. The explicit dependence of $R_D$ and $R_{D^*}$ on the parameters $\alpha_{23}$,
which determines the mass of the charged Higgs $M_{H^\pm}$, $\theta_{13}$, and the phase $\alpha$,
can be understood through the following approximate analytical expressions:
\begin{equation}
R_D=0.298+\Gamma_{l,\nu}^{H^\pm} \sin (\theta_{13}) \left(\frac{8.854\times 10^{-4}
\Gamma_{l,\nu}^{H^\pm} \sin (\theta_{13})}{(\alpha_{23})^2}-2.39\times 10^{-2}
\Re\left(\frac{e^{-i \alpha }}{\alpha_{23}}\right)\right)
\end{equation}
\begin{equation}
R_D^*=0.254+\Gamma_{l,\nu}^{H^\pm} \sin (\theta_{13}) \left(\frac{2.936\times 10^{-5}
\Gamma_{l,\nu}^{H^\pm} \sin (\theta_{13})}{(\alpha_{23})^2}+1.496\times 10^{-3}
\Re\left(\frac{e^{-i \alpha }}{\alpha_{23}}\right)\right)
\end{equation}
We can obtain the regions in the $(\theta^R_{13},\alpha)$
parameter space in which the anomalies are satisfied through
varying $\theta^R_{13}$ and $\alpha$ while assigning fixed values
of the other parameters. As an example, we take the fixed values
$t_\beta\simeq 0.2074$ and $\alpha_{32} = 0.00416$ which result in
$m_{H^\pm} = 305$ GeV and the other parameters are fixed as
before. In Fig.\ref{ZBs11}, we show the experimentally allowed
$2\sigma$ region in the $(\theta^R_{13},\alpha)$ plane satisfying
$R_{D^*}$ and $R_D$ together in red color. In the same plot, the
regions in green color are the allowed regions, in the
$(\theta^R_{13},\alpha)$ plane for the set of the input parameters
we use, by all constraints discussed above in the previous
section. Clearly, the imposed constraints have a sensible effect
on the parameter space as large part of this space is excluded by
the constraints. Moreover, as shown in the plot, their is a small
region in the parameter space which is allowed by all the
aforementioned constraints in which $R_{D^*}$ and $R_D$ anomalies
are satisfied together. Specifically, this region is the
intersection region of the two colored regions in the plot. It
should be noted that this conclusion corresponds to our particular
example of the input parameters stated above. Taking other values
of the input parameters $t_\beta$, $\alpha_{32}$ and $v_R$ may
lead to another regions in $(\theta^R_{13},\alpha)$ plane that
respect all constraints and satisfy $R_{D^*}$ and $R_D$ anomalies
together. For instances, in table \ref{bech}, we list several
benchmarks obtained upon variation of $\alpha_{32}$ while keeping
other input parameters as before.
In Fig. \ref{correlation}, we present the correlation between
$R_D$ and $R_{D^*}$ for the same set of the parameter space
considered in the scan over the values and ranges mentioned in the
beginning of this section. Only points that satisfy all the bounds
discussed in section \ref{higgs} and the $2\sigma$ range of
$F_L(D^*)$ and $P_{\tau}(D^*)$ are included here. It is remarkable
from this figure that both $R_D$ and $R_{D^*}$ are satisfied for a
lot of points in the parameter space, thanks to the complex mixing
of right-handed quarks.

It should be noted that, if the $t b H^+$ Yukawa coupling is
non-negligible ($c_{\theta13}^R$ is non-zero), there will be
non-negligible $tt \phi$ couplings thanks to the SU(2) invariance
where $\phi$ stands for the neutral extra scalars: $H$ and $A$. As
a consequence, $gg \to \phi \to \tau \tau$ would also impose
constraints on  the parameter space. To consider this possibility,
we have analyzed the processes $gg\rightarrow \phi\rightarrow
\tau\tau$. We have found that the corresponding cross-sections are
less than $10^{-8}$ Pb, indicating that they do not impose any
significant constraints on our analysis.

We turn now to the discussion of our predictions of the $D^\ast$
and $\tau$ longitudinal polarizations and tau polarization
observable $P_\tau (D)$. For this purpose we list in table
\ref{bech} a selective set of  6 benchmarks representing part of
our results. Recall that  the corresponding measurements, reported
by Belle collaborations, are $F^{Expt}_L(D^\ast) = 0.60 \pm 0.08
\pm 0.035$~\cite{Belle:2019ewo} and $P^{Expt}_\tau(D^\ast) = -
0.38 \pm
0.51^{+0.21}_{-0.16}$~\cite{Hirose:2017dxl,Hirose:2016wfn,Adamczyk:2019wyt}.
Clearly, the uncertainty in the measurement of
$P^{Expt}_\tau(D^\ast)$ is so large and thus we only restrict
ourselves to  $F_L(D^\ast)$.   The SM prediction is estimated to
be $F^{\rm SM}_L(D^\ast)= 0.464 \pm 0.010$~\cite{Bordone:2019vic}.
Clearly from table \ref{bech}, the new contributions of the model
under concern can increase the central value of the SM by about
$11\%$ for the listed set of the benchmarks. We also noted there
are  points in our scan that imply a larger enhancement which
makes our predictions closer to the experimentally measured
central value. On the other hand, the enhancement in the tau
polarization $P_\tau (D)$ can reach 6 times its SM predicted value
as can be noted from the same table. This can serve as test of the
model under consideration once this observable is measured.

\begin{table}[ht]
\begin{center}
\scalebox{0.6}{%
\begin{tabular}{|c|c|c|c|c|c|c|c|c|c|c|c|c|c|c|c|c|}
\hline
\textbf{Parameter} & $\Gamma^{H^\pm}_{l,\nu}$ & $\theta^R_{13}$ & $\alpha$& $\alpha_{32}$ & $m_H^{\pm}$ &
$\Delta M_{B_d}$ & $\Delta M_{B_s}$ &$ BR(B_c^- \to \tau^- \bar{\nu}_{\tau})$ & $R_D$ & $R_{D^*}$  &$F_L(D^*)$ & $P_{\tau}(D^*)$&$P_{\tau}(D)$&$B^0_s\to \mu^{+} \mu^{-}$&$|C^{H^\pm}_7+0.19\,C^{H^\pm}_8 |$ \\
\hline
\textbf{BP1} & 0.03 & 1.5 & 2.4 &0.00186& 204 & 0.6 & 1.0 & $50\%$  & 0.296 & 0.285  & 0.51 & -0.3&2.1& $3.56\times 10^{-9}$ & 0.027 \\
\textbf{BP2} & 0.03 & 1.7 & 2.2 &0.00216& 220 & 0.7 & 1.0 & $40\%$ & 0.31 & 0.27  & 0.5 & -0.37 &1.8& $4.1\times 10^{-9}$& 0.02  \\
\textbf{BP3} & 0.08 & 2.55 & 2.0 &0.00366& 286 & 0.94 & 1 & $37\%$ & 0.373 & 0.274  & 0.49 & -0.4&1.3& $3.89\times 10^{-9}$ & 0.019  \\
\textbf{BP4} & 0.18 & 2.1 & 2.3 & 0.00476 &327 & 0.89 & 0.99 & $42\%$ & 0.33 & 0.283  & 0.50 & -0.33 &1.7& $3.72\times 10^{-9}$& 0.027 \\
\textbf{BP5} & 0.16 & 2.3 & 2.2 &0.00616& 371 & 0.93 & 0.996 & $40\%$  & 0.35 & 0.284  & 0.51 & -0.347 &1.0& $3.98\times 10^{-9}$& 0.019 \\
\textbf{BP6} & 0.14 & 2.3 & 2.1 &0.00716 & 400 & 0.99 & 0.97 & $34\%$ & 0.32 & 0.27  & 0.48 & -0.4 &1.4& $3.54\times 10^{-9}$& 0.027  \\
\hline
\end{tabular}}
\caption{\label{Bnchmrkpnts} A set of selective benchmarks of the
parameters  and intended predictions resulting from the scan after
taking into account the constraints discussed in the previous
section and requirements stated in the beginning of this section.
\label{bech} }
\end{center}
\end{table}

\begin{table}[ht]
\begin{center}
\scalebox{0.6}{%
\begin{tabular}{|c|c|c|c|c|c|c|c|}
\hline
\textbf{Parameter} & $\Gamma^{H^\pm}_{l,\nu}$ & $\theta^R_{13}$ & $\alpha$& $\alpha_{32}$ & $m_H^{\pm}$ &
$Re(C^{NP}_{9\mu})/C^{SM}_{9\mu}$ & $Re(C^{NP}_{10\mu})/C^{SM}_{10\mu}$  \\
\hline
\textbf{BP1} & 0.03 & 1.5 & 2.4 &0.00186 & 204 & 0.07 & 0.02 \\
\textbf{BP2} & 0.03 & 1.7 & 2.2 &0.00216 & 220 & 0.08 &0.02 \\
\textbf{BP3} & 0.08 & 2.55 & 2.0 &0.00366& 286 &-0.039  & -0.01 \\
\textbf{BP4} & 0.18 & 2.1 & 2.3 & 0.00476 &327  &0.04  &0.01 \\
\textbf{BP5} & 0.16 & 2.3 & 2.2 &0.00616& 371 & 0.01  & 0.017  \\
\textbf{BP6} & 0.14 & 2.3 & 2.1 &0.00716 & 400 & 0.03 &0.009  \\
\hline
\end{tabular}}
\caption{ $Re(C^{NP}_{9\mu})/C^{SM}_{9\mu}$ and
$Re(C^{NP}_{10\mu})/C^{SM}_{10\mu}$ for the set of benchmark
points considered in table \ref{bech}.  \label{bechc9} }
\end{center}
\end{table}

Searching for new physics via  $B$ meson  exclusive decays
originating at the quark level from the $b\to s\ell^{+}\ell^{-}$
transitions has gained a lot of attention in the last decades. In
the framework of the SM, these transitions take place only at
loop-level. Consequently, their amplitudes are suppressed, for
instances, by factors accounting for the integration over the
momenta running in the loop. This in turn leads to deviations of
the measured branching ratios of $B\to K\mu^{+}\mu^{-}$, $B\to
K^{\ast}\mu^{+}\mu^{-}$, and $B_{s}\to\phi\mu^{+}\mu^{-}$ decays.
from their SM predictions. Also, the angular observable
$P_{5}^{\prime}$ in the $B^{0}\to K^{\ast 0}\mu^{+}\mu^{-}$ decay,
\cite{Descotes-Genon:2012isb,Descotes-Genon:2013vna}, has shown
tension with the SM values. For instance, ATLAS
\cite{ATLAS:2018gqc}, and LHCb \cite{LHCb:2015svh,LHCb:2020lmf},
measured the value of $P_{5}^{\prime}$ in the kinematical region
$4.0<q^{2}<6.0$ $\text{GeV}^{2}$ and found departure from the SM
value to be more than $3\sigma$ \cite{Aebischer:2018iyb}.
Furthermore Belle \cite{Belle:2016xuo,Belle:2016fev} and CMS
\cite{CMS:2017rzx} measured the value of $P_{5}^{\prime}$ for the
same decay mode in $q^{2}$ bin $4.0<q^{2}<8.0$ $\text{GeV}^{2}$
and $6.0<q^{2}<8.68$ $\text{GeV}^{2}$ respectively. Belle
measurement shows the deviation of $2.6\sigma$ from the SM
prediction and CMS measurement shows a discrimination of $1\sigma$
from the SM value. Their is a possibility that the tensions
between the measurements and the SM predictions of all these
observables can be relaxed if one considers NP models affecting
the $b\to s\ell^{+}\ell^{-}$ transitions. In the literature,
various performed model independent global fit analyses
\cite{Alguero:2021anc,Descotes-Genon:2015uva,Altmannshofer:2017fio,Alok:2017sui,Altmannshofer:2017yso,
Geng:2017svp,Ciuchini:2017mik,Capdevila:2017bsm,Alguero:2019ptt,Alok:2019ufo,Ciuchini:2019usw,Datta:2019zca,Aebischer:2019mlg,
Kowalska:2019ley,Arbey:2019duh,Bhattacharya:2019dot,Biswas:2020uaq,Alok:2022pjb}
based on the assumption that NP present only in the muon sector
revealed that two simple one-dimensional (1D) NP scenarios (S1)
$C_{9\mu}^{\text{NP}}$ or (S2)
$C_{9\mu}^{\text{NP}}=-C_{10\mu}^{\text{NP}}$, that give better
fit to all the data, with preferences reaching $\approx5-6\sigma$
compared to the SM \cite{Zaki:2023nlw}.

The observable $P '_5$ is sensitive to the real parts of
$C_{9\mu}^{\text{NP}}$ and $C_{10\mu}^{\text{NP}}$. On the other
hand, the ratios $R_{K^{(\ast)}}$, defined as
$R_{K^{(\ast)}}=\frac{\mathcal{B}(B\to
K^{(\ast)}\mu^{+}\mu^{-})}{\mathcal{B}(B\to
K^{(\ast)}e^{+}e^{-})}$, are also sensitive to
$C_{9\mu}^{\text{NP}}$ and $C_{10\mu}^{\text{NP}}$. Recently the
updated measurements of  $R_{K^{(\ast)}}$,
\cite{LHCb:2022qnv,LHCb:2022zom} have put stringent constraints on
the NP couplings and the NP models. This in turn lead to a strong
constraints on the Wilson coefficients $C_{9\mu}^{\text{NP}}$ and
$C_{10\mu}^{\text{NP}}$ which affect the aforementioned
observables. In the model under concern we found that it is not
possible to accommodate $P '_5$ and other observables related to
the $b\to s\ell^{+}\ell^{-}$ transitions through
$Re(C^{NP}_{9\mu})$ and $Re(C^{NP}_{10\mu})$. To make it more
clear, we show in table \ref{bechc9}
$Re(C^{NP}_{9\mu})/C^{SM}_{9\mu}$ and
$Re(C^{NP}_{10\mu})/C^{SM}_{10\mu}$ for the set of benchmark
points in table \ref{bech}. As can be seen from the table there is
no sizable enhancement due to the new contributions of the model
under consideration compared to the corresponding SM Wilson
coefficients.

\section{Conclusion \label{sec:conclusion}}

In this work we have explored the possibility of resolving the
tension between the SM prediction and the experimental results of
the  $R_D$ and $R_{D^*}$ ratios using a low scale left-right
symmetric model based on $SU(3)_C\times SU(2)_L\times
SU(2)_R\times U(1)_{B-L}$.  The scalar sector of the model
contains charged Higgs boson with masses that can be chosen in the
order of hundreds GeV without any conflict with direct search
constraints. We have shown that integrating out the charged Higgs
mediating the tree-level diagrams  generates a set of non
vanishing scalar Wilson coefficients contributing to the effective
Hamiltonian governing the transition $b\to c\tau\bar\nu$ and hence
to the ratios $R_{D^*}$ and $R_D$ and the $D^\ast$ and $\tau$
polarizations.

 The dependency of the scalar Wilson coefficients on the matrix
 elements of the quark mixing angle in the right sector turns to
 be important. We emphasized that the mixing element $(V^R_{CKM})_{23}$
 should be complex in order to satisfy both $R_D$ and
 $R_{D^*}$. We have also shown the complex phase associated with this mixing element
 is essential to accommodate the experimental results of the
 ratios for charged Higgs masses of order 300 GeV  while respecting the
constraints from ${\rm BR}(B^-_c \to \tau^- \bar{\nu}_\tau) $,
$B_{s(d)}-\bar B_{s(d)}$ mixing and other relevant constraints
discussed above.

%%%%%%%%%%%%%%%%%%%%%%%%%%
\section*{acknowledgements}
The work of K. E. and S. K. is partially supported by Science, Technology $\&$ Innovation Funding Authority (STDF) under grant number 37272.
%%%%%%%%%%%%%%%%%%%%%%%%%%

\appendix

\section{\label{app}}

\subsection{Wilaon coefficients relevant to the process $B_s \to \ell_{A}^{+} \ell_{A}^{-}$\label{Bmumu}}
Following Ref. \cite{Crivellin:2013wna}, the neutral Higgs
non-vanishing  Wilson coefficients can be expressed as
\begin{equation}
\renewcommand{\arraystretch}{1.6}
\begin{array}{l}
  C_S^{b s}= \dfrac{{ \pi^{2}}}{2 G_{F}^{2}
    M_{W}^{2}}\sum\limits_{k = 1}^3 \dfrac{1}{m^{2}_{H_{k}^{0}}}\,
  \left( \Gamma _{\ell_{A} \ell_{A}}^{LR\, H_{k}^{0}}  + \Gamma
  _{\ell_{A} \ell_{A}}^{RL \, H_{k}^{0}}          \right) \,  \Gamma
  _{b s}^{RL\, H_{k}^{0}}        \\

  C_P^{b s}  = \dfrac{{ \pi^{2}}}{2 G_{F}^{2}
    M_{W}^{2}}\sum\limits_{k = 1}^3 \dfrac{1}{m^{2}_{H_{k}^{0}}}\,
  \left( \Gamma _{\ell_{A} \ell_{A}}^{LR\, H_{k}^{0}}  - \Gamma
  _{\ell_{A} \ell_{A}}^{RL \, H_{k}^{0}}          \right) \,  \Gamma
  _{b s}^{RL\, H_{k}^{0}}        \\

  C_S^{' b s}= \dfrac{{ \pi^{2}}}{2 G_{F}^{2}
    M_{W}^{2}}\sum\limits_{k = 1}^3 \dfrac{1}{m^{2}_{H_{k}^{0}}}\,
  \left( \Gamma _{\ell_{A} \ell_{A}}^{LR\, H_{k}^{0}}  + \Gamma
  _{\ell_{A} \ell_{A}}^{RL \, H_{k}^{0}}          \right) \,  \Gamma
  _{ b s}^{LR\, H_{k}^{0}}        \\

   C_P^{' b s} = \dfrac{{ \pi^{2}}}{2 G_{F}^{2}
     M_{W}^{2}}\sum\limits_{k = 1}^3 \dfrac{1}{m^{2}_{H_{k}^{0}}}\,
   \left( \Gamma _{\ell_{A} \ell_{A}}^{LR\, H_{k}^{0}}  - \Gamma
   _{\ell_{A} \ell_{A}}^{RL \, H_{k}^{0}}          \right) \,  \Gamma
   _{ b s}^{LR\, H_{k}^{0}}           \, .
 \end{array} \label{WCN}
\end{equation}

with vertex

\begin{equation}
{\Gamma_{b s}^{RL\, A^0 } }= - \frac{1}{\sqrt{2}}
\Big(\sum_{b=1}^{3}(V^R_{CKM})^*_{3 b}
\sum_{a=1}^{3}(V^L_{CKM})^*_{2 a} y^{Q^*}_{a b}   Z_{{3 2}}^{A}  +
\sum_{b=1}^{3}(V^R_{CKM})^*_{3 b} \sum_{a=1}^{3}(V^L_{CKM})^*_{2
a} \tilde{y}^{Q^*}_{a b}   Z_{{3 1}}^{A} \Big)
\end{equation}

\begin{align}
{\Gamma_{ b s}^{LR \, A^0 } }= \frac{1}{\sqrt{2}}
\Big(\sum_{b=1}^{3}\sum_{a=1}^{3}(V^L_{CKM})_{3 a} y^{Q}_{a b}
(V^R_{CKM})_{2 b}  Z_{{3 2}}^{A}  +
\sum_{b=1}^{3}\sum_{a=1}^{3}(V^L_{CKM})_{3 a} \tilde{y}^{Q}_{a b}
(V^R_{CKM})_{2 b}  Z_{{3 1}}^{A} \Big)
\end{align}

\begin{equation}
{\Gamma_{b s}^{RL \,h^k} }= i \frac{1}{\sqrt{2}} \Big(-
\sum_{b=1}^{3}(V^R_{CKM})^*_{3 b} \sum_{a=1}^{3}(V^L_{CKM})^*_{2
a} y^{Q^*}_{a b}   Z_{{k 2}}^{H}  + \sum_{b=1}^{3}(V^R_{CKM})^*_{3
b} \sum_{a=1}^{3}(V^L_{CKM})^*_{2 a} \tilde{y}^{Q^*}_{a b}   Z_{{k
1}}^{H} \Big)
\end{equation}

\begin{align}
{\Gamma_{b s}^{LR \,h^k } }= i \frac{1}{\sqrt{2}} \Big(-
\sum_{b=1}^{3}\sum_{a=1}^{3}(V^L_{CKM})_{3 a} y^{Q}_{a b}
(V^R_{CKM})_{2 b}  Z_{{k 2}}^{H}  +
\sum_{b=1}^{3}\sum_{a=1}^{3}(V^L_{CKM})_{3 a} \tilde{y}^{Q}_{a b}
(V^R_{CKM})_{2 b}  Z_{{k 1}}^{H} \Big)
\end{align}

\begin{equation}
{\Gamma_{\ell_A, \ell_A}^{RL\,A^0 } }= \frac{1}{\sqrt{2}} \Big(-
y^{L^*}_{A A} Z_{{3 2}}^{A} + \tilde{y}^{L^*}_{A A} Z_{{3 1}}^{A}
\Big)
\end{equation}

\begin{align}
{\Gamma_{\ell_A \ell_A}^{LR \,A^0 } }= - \frac{1}{\sqrt{2}} \Big(-
y^{L}_{A A} Z_{{3 2}}^{A}  + \tilde{y}^{L}_{A A} Z_{{3 1}}^{A}
\Big)
\end{align}

\begin{equation}
{\Gamma_{\ell_A \ell_A}^{RL\,h^k } }= -i \frac{1}{\sqrt{2}} \Big(
y^{L^*}_{A A}   Z_{{k 2}}^{H}  + \tilde{y}^{L^*}_{A A}   Z_{{k
1}}^{H} \Big)
\end{equation}

\begin{align}
{\Gamma_{\ell_A \ell_A}^{LR\, h^k } }= -i \frac{1}{\sqrt{2}} \Big(
y^{L}_{A A} Z_{{k 2}}^{H}  + \tilde{y}^{L}_{A A} Z_{{k 1}}^{H}
\Big)
\end{align}
The expressions of $Z_{{ij}}^{A,H}$  can be found in
Ref.\cite{Ezzat:2021bzs}. The Wilson coefficients at low energy
scale $\mu_{{low}}, $ $C_{S,P}^{(\prime)bs}(\mu_{{low}})$, can be
obtaining from their corresponding ones listed in Eq.(\ref{WCN})
using the relation \cite{Crivellin:2013wna}

\begin{eqnarray}
C_{S,P}^{(\prime)bs}(\mu_{{low}})=\frac{m_{q}(\mu_{{low}})}{m_{q}(\mu_{{high}})}
\, C_{S,P}^{(\prime)bs}(\mu_{{high}})\,,
\end{eqnarray}
here $m_q$ represents the running quark mass with the appropriate
number of active flavors. Finally, it should be noted that since
the Wilson coefficients are given at the matching scale, $m_b$ and
$m_t$ must be evaluated at this scale \cite{Crivellin:2013wna}.

\subsubsection{{ $B_{q}-\bar B_{q}$ mixing\label{Bmix} }}

The contributions of the NP to the effective hamiltonian
generating $\Delta B=2$ transitions, q = d,s, can be written as

\be \mathcal{H}^{\Delta Q=2}_\mathrm{eff} = -\frac{4
G_F}{\sqrt{2}} |V_{tb}V_{tq}^*|^2 \bigg\{\sum^5_{i=1} C^{NP}_i
Q_i+\sum^3_{i=1} \tilde C^{NP}_i \tilde Q_i \bigg\}\,.
\label{BsH}\ee The four-quark operators are given as \bea Q_1 &=&
\left(\bar{q}\gamma^{\mu} P_L b\right)
\left(\bar{q}\gamma_{\mu} P_L b\right)\nonumber\\
Q_2 &=& ({\bar q}_{\alpha} P_{L} b_{\alpha})\, ({\bar q}_{\beta}
P_{L} b_{\beta}) \nonumber\\
Q_3 &=& ({\bar q}_{\alpha} P_{L} b_{\beta})\, ({\bar q}_{\beta}
P_{L} b_{\alpha}) \nonumber\\
Q_4 &=& ({\bar q}_{\alpha} P_{L} b_{\alpha})\, ({\bar q}_{\beta}
P_{R} b_{\beta})
\nonumber\\
Q_5 &=& \left(\bar{q}_{\alpha} P_L b_{\beta}\right)
\left(\bar{q}_{\beta} P_R b_{\alpha}\right)\label{mixH}\eea The
operators $\tilde Q_{1,2,3}$ can be obtained from the operators
$Q_{1,2,3}$ by the replacement $L \leftrightarrow R$. At
tree-level, we have contributions to $\mathcal{H}^{\Delta
Q=2}_\mathrm{eff}$ from only neutral Higgs bosons mediation.
Denoting these contributions by $C_i^{H_{k}^{0}}$, we find that
their expressions can be written as
\bea C_2^{H_{k}^{0}}(\mu_{H_{k}^{0}}) &=& \frac {\sqrt{2}}{4 G_F
|V_{tb}V_{tq}^*|^2}
 \sum\limits_{k = 1}^3  \, \dfrac{  1}{2
m_{H_{k}^{0}}^{2}} \, (\Gamma _{\;b q}^{LR\,H_{k}^{0}\star})^{2}
\\ \nonumber
 C_2^{\prime\,H_{k}^{0}} (\mu_{H_{k}^{0}}) &=& \frac {\sqrt{2}}{4 G_F |V_{tb}V_{tq}^*|^2}\sum\limits_{k = 1}^3  \, \dfrac{  1}{2 m_{H_{k}^{0}}^{2}} \,
(\Gamma _{\;q b}^{LR\,H_{k}^{0}})^{2}  \\ \nonumber
C_4^{\,H_{k}^{0}} (\mu_{H_{k}^{0}}) &=& \frac {\sqrt{2}}{4 G_F
|V_{tb}V_{tq}^*|^2} \sum\limits_{k = 1}^3  \, \dfrac{
1}{m_{H_{k}^{0}}^{2}} \, \Gamma _{\;q b}^{LR\,H_{k}^{0}} \, \Gamma
_{\;b q}^{LR\,H_{k}^{0}\star} \label{Wnhn}\eea
The quantities $\Gamma _{\;q b}^{AB\,H_{k}^{0}}$ are defined as
before. We remark from the above expressions that, the neutral
Higgs bosons contributes only to the scalar color singlet
four-quark operators in Eq.(\ref{mixH}). In
Ref.\cite{DiLuzio:2019jyq}, it was pointed out that the
contributions of the operators that contains scalar and tensor
Dirac structures are highly disfavored by the fits to $b \to s $
data and hence their contributions can be neglected. The charged
Higgs contributions to $\mathcal{H}^{\Delta Q=2}_\mathrm{eff}$
originate at one loop-level and can be expressed as\bea
C^{H^\pm}_{1}(\mu_{H^\pm})&=&-\frac{1}{256\sqrt{2} \pi^2 G_F
m_{H^+}^2 (V_{tb}V_{t q}^*)^2} \sum_{k,\ell} \Gamma_{\ell b
}^{H^\pm\,RL\,\rm{eff}} (\Gamma_{k q }^{H^\pm\,RL\,\rm{eff}})^*
\bigg[ \Gamma_{k b }^{H^\pm\,RL\,\rm{eff}} (\Gamma_{\ell q
}^{H^\pm\,RL\,\rm{eff}})^*
 G_1(x_k,x_\ell)\nonumber \\
 &-&\frac{4g^2m_{u_k} m_{u_\ell} }{m_{H^+}^2} V_{k b}V^*_{\ell q}G_2(x_k,x_\ell,x_W)
 +\frac{g^2 m_{u_k}m_{u_\ell}}{m_W^2} V_{k b} V^*_{\ell q}G_3(x_k,x_\ell,x_W)
 \bigg],
\label{WcH}\eea where the loop functions $G_i$ for $i=1,2,3$ are
given as
\begin{align}
  G_1(x,y) &=\frac{1}{x-y}\left[
    \frac{x^2\log x}{(1-x)^2}+\frac{1}{1-x}-
    \frac{y^2\log y}{(1-y)^2}-\frac{1}{1-y}
    \right],
\label{G1_func}
  \\
  G_2(x,y,z) &=-\frac{1}{(x-y)(1-z)}
  \left[
    \frac{x \log x}{1-x}-\frac{y\log y}{1-y}-\frac{x\log\frac{x}{z}}{z-x}
    +\frac{y\log \frac{y}{z}}{z-y}
    \right],
  \\
  G_3(x,y,z) &=-\frac{1}{x-y}
  \left[
    \frac{1}{1-z}\left(\frac{x\log x}{1-x}-\frac{y\log y}{1-y}\right)
%    \right.
%    \nonumber \\
%    &\left.
    -\frac{z}{1-z}\left(
    \frac{x\log \frac{x}{z}}{z-x}-\frac{y\log \frac{y}{z}}{z-y}\right)\right].
\end{align}
and $x_W=m_W^2/m_{H^+}^2$ and $x_k=m_{u_k}^2/m_{H^+}^2$.  The next
step is to calculate the matrix elements of the operators $Q_i$ at
the scale $\mu=\mu_b$ and to run the Wilson coefficients from the
electroweak scale to the scale $\mu=\mu_b$. The contribution to
the $B_{q}-\bar B_{q}$ mixing amplitudes induced by a given NP
scale coefficient $C_i(\mu=\mu_{NP})$, denoted by $\langle B_q |
\mathcal{H}^{\Delta B=2}_\mathrm{eff} |\bar B_q\rangle_i$, as a
function of $\alpha_s(\mu_{NP})$ and the scale $\mu=\mu_b$ is
given as \cite{UTfit:2007eik} (see also \cite{Becirevic:2001jj})

\begin{equation}\label{eq:running}
\langle B_q | \mathcal{H}^{\Delta B=2}_\mathrm{eff} |\bar
B_q\rangle_i= \sum_{j=1}^5 \sum_{r=1}^5\, \left(b_j^{(r,i)}+
\eta\, c_j^{(r,i)}\right)\,\eta^{a_j}\, C_i(\mu_{NP})\,B_i^{B_q}\,
\langle B_s | Q_r | \bar B_q\rangle\,,
\end{equation}
where $\eta = \alpha_s(\mu_{NP}))/\alpha_s(m_t)$, $a_j,
b_j^{(r,i)}$ and $ c_j^{(r,i)}$ are magic numbers given in
\cite{UTfit:2007eik} and $B_{B_s}^i$ are the $B$ parameters that
can be found in Table 9 in Ref.\cite{Bauer:2015fxa}.  It should be
noted that the magic numbers for the evolution of the Wilson
coefficients $\tilde C_{1-3}$ are the same as the ones for the
evolution of $C_{1-3}$ \cite{Becirevic:2001jj}. Moreover, in the
basis $\tilde Q_{i}$, the $\,B_i^{B_q}$ parameters  and the
hadronic matrix elements $ \langle B_q | \tilde Q_r | \bar
B_q\rangle$ are equal to their corresponding ones in the basis
$Q_{i}$ \cite{Crivellin:2021lix}. The matrix elements are given by

\begin{align}\label{eq:matrixelements}
\langle  B_q | Q_1 | \bar B_q\rangle&=\frac{1}{3}M_{B_q}\, f_{B_q}^2\,, \nonumber\\
\langle  B_q | Q_r | \bar
B_q\rangle&=N_r\,\left(\frac{M_{B_q}}{m_q+m_b}\right)^2\, M_{B_q}
f_{B_q}^2\,,
\end{align}
with $N_r= (-5/24, 1/24, 1/4, 1/12)$ for $r=(2,3,4,5)$. With all
this in hand, it is direct to calculate the quantity $ \Delta
M_{q}$ \cite{Buchalla:1995vs,Buras:1998raa} and thus derive the
bounds on the parameter space using the measured value of $ \Delta
M_{q}$.

 The $K-\bar K$ mixing can be studied in a similar way used
 above in the $B_q-\bar B_q$ mixing. In fact,  it is possible to have
a non-vanishing phase in the second column of the
$V^R_{\text{CKM}}$ matrix, which turns out to be irrelevant and
has no effect on the $R_D$ and $R_D^*$ result. However, this
imaginary phase can affect the bound from other flavor observables
e.g. $\epsilon_K$. In particular, through the contributions from
the neutral Higgs mediating the tree-level diagrams. In order to
provide the ingredient required for the estimation of
$\epsilon_K$, we define
\begin{equation}\label{eq:epsK}
C_{\epsilon_K}=\frac{Im \,\langle  K^0| {\cal
H}_\text{full}^{\Delta S=2}| \bar K^0 \rangle }{Im \,\langle K^0|
\mathcal{H}_\mathrm{SM}^{\Delta S=2}| \bar K^0 \rangle }\,,
\end{equation}
where ${\cal H}_\text{full}^{\Delta S=2} =
\mathcal{H}_\mathrm{SM}^{\Delta S=2} +{\cal H}_\text{NP}^{\Delta
S=2} $. The calculation of $\langle K^0| {\cal
H}_\text{full}^{\Delta S=2}| \bar K^0 \rangle_i$ can be done in a
similar manner to $\langle B_q | \mathcal{H}^{\Delta
B=2}_\mathrm{eff} |\bar B_q\rangle_i$ with the replacement of $bs
\leftrightarrow s d$ in the Wilson coefficients listed above. The
$K-\bar K$ mixing can be evaluated using \cite{UTfit:2007eik}
\begin{equation}\label{eq:running}
\langle \bar K | \mathcal{H}^{\Delta S=2}_\mathrm{eff} |
K\rangle_i= \sum_{j=1}^5 \sum_{r=1}^5\, \left(b_j^{(r,i)}+ \eta\,
c_j^{(r,i)}\right)\,\eta^{a_j}\, C_i^{sd}(\mu)\,B_i^K\, \langle
\bar K | Q_r^{sd} | K\rangle\,,
\end{equation}
as before  $a_i, b_j^{(r,i)}$ and $ c_j^{(r,i)}$ are "magic
numbers" listed in \cite{Ciuchini:1998ix} and $B_K^i$ are the $B$
parameters collected in Table 8. The matrix elements are given by
\begin{align}\label{eq:matrixelements}
\langle \bar K | Q_1^{sd} | K\rangle&=\frac{1}{3}M_K\, f_k^2\,, \nonumber\\
\langle  \bar K | Q_r^{sd} |
K\rangle&=N_r\,\left(\frac{M_K}{m_d+m_s}\right)^2\, M_K f_k^2\,,
\end{align}
with $N_r= (-5/24, 1/24, 1/4, 1/12)$ for $r=(2,3,4,5)$. In the
next step, the Wilson coefficients are evolved down from the mass
scale of the Higgs scalars to the scale $\mu= 2$ GeV  at which the
hadronic matrix elements are evaluated using the RG equations in
\cite{Ciuchini:1998ix}.  Doing so, it is quite forward and direct
to compute the constraint on the non-vanishing phase in the second
column of the $V^R_{\text{CKM}}$ matrix using the bound from
$\epsilon_K$.

\end{document}